\documentclass[journal]{IEEEtran}
\usepackage{cite}
\usepackage{babel}
\usepackage{amsmath,amssymb,amsfonts,amsthm, bm}
\usepackage{siunitx}
\usepackage{graphicx}
\usepackage{textcomp}
\usepackage{enumerate}
\usepackage[capitalise]{cleveref}
\usepackage{color}
\usepackage{algorithm,algpseudocode}
\usepackage{tikz}
\usepackage{textcomp}
\usepackage[acronym]{glossaries}

\def\s{{\mathbf s}}
\def\u{{\mathbf u}}
\def\e{{\mathbf e}}
\def\g{{\mathbf g}}

\def\v{{\mathbf v}}
\def\x{{\mathbf x}}
\def\y{{\mathbf y}}
\def\z{{\mathbf z}}
\def\h{{\mathbf h}}
\def\w{{\mathbf w}}

\def\Null{{\mathbf 0}}

\def\H{{\mathbf H}}

\def\X{{\mathbf X}}

\def\Y{{\mathbf Y}}

\def\I{{\mathbf I}}

\def\setA{\mathcal{A}}
\def\setB{\mathcal{B}}
\def\setC{\mathcal{C}}
\def\setD{\mathcal{D}}
\def\setE{\mathcal{E}}

\def\setH{\mathcal{H}}
\def\setI{\mathcal{I}}
\def\setJ{\mathcal{J}}
\def\setK{\mathcal{K}}

\def\setN{\mathcal{N}}
\def\setP{\mathcal{P}}
\def\setQ{\mathcal{Q}}
\def\setR{\mathcal{R}}
\def\setS{\mathcal{S}}

\def\setU{\mathcal{U}}
\def\setV{\mathcal{V}}
\def\setW{\mathcal{W}}
\def\setX{\mathcal{X}}

\def\bigo{\mathcal{O}}

\def\RR{{\mathbb R}}
\def\CC{{\mathbb C}}
\def\NN{{\mathbb N}}

\def\minimize{\mathrm{minimize}}
\def\argmin{\mathrm{arg\ min}}
\def\argmax{\mathrm{arg\ max}}
\def\find{\mathrm{find}}

\def\diag{\mathrm{diag}}

\def\prox{{\mathrm{prox}}}
\def\sign{\mathrm{sign}}
\def\lev{{\mathrm{lev}}}
\def\Id{I}

\def\T[#1]{T_{#1}}

\def\Xk[#1]{{\X_{#1}}}
\def\Yk[#1]{{\Y_{#1}}}

\def\normbarspacing{-0.25ex} % controls the spacing between the 3 norm bars
\newcommand{\vertiii}[1]{{\left\vert\kern\normbarspacing\left\vert\kern\normbarspacing\left\vert #1 
		\right\vert\kern\normbarspacing\right\vert\kern\normbarspacing\right\vert}}
\def\Norm[#1]{\vertiii{#1}}
\def\relu[#1]{\left(#1\right)_+}

\def\fix{\mathrm{Fix}}

\usepackage[framemethod=tikz]{mdframed}

\newtheorem{theorem}{Theorem}
\newtheorem{prop}{Proposition}

\newtheorem{lemma}{Lemma}

\newtheorem{fact}{Fact}

\theoremstyle{definition}
\newtheorem{definition}{Definition}
\theoremstyle{remark}

\newtheorem{comment}{Commment}
\newtheorem{remark}{Remark}

\renewcommand{\triangleq}{:=}

\def\Nt{K}
\def\Nr{N}

\def\xml{\s^\star}

\def\shrink{\phi}
\def\rhoML{\rho^\star}
\def\rhohat{\hat\rho}
\def\nmin{n_0}
\def\fltwo{f_{{\ell_2}}}
\def\flone{f_{{\ell_1}}}
\def\vlone{\v_n^{(\ell_1)}}
\def\vltwo{\v_n^{(\ell_2)}}

\def\apsm{\texttt{APSM}}
\def\apsmlone{\texttt{APSM-L1}}
\def\apsmltwo{\texttt{APSM-L2}}
\def\amp{\texttt{IO-LAMA}}
\def\oamp{\texttt{OAMP}}
\def\lmmse{\texttt{LMMSE}}

\def\boxdetector{\texttt{Box Detector}}

\def\nmax{n_{\max}}

\def\emin{\varepsilon_1}
\def\emax{\varepsilon_2}
\def\phin{\bm{\Phi}_n}

\def\fix{\mathrm{Fix}}
\def\lplusone{\ell^1_+(\NN)}
\def\vhat{\hat\v}

\def\omg{{\bm{\omega}}}

\def\minimize{\mathrm{minimize}}

\def\wlone[#1]{{\|#1\|_1^\omg}}
\def\wloneN[#1]{{\|#1\|_1^{\omg^{(n)}}}}

\def\Null{\mathbf{0}}

\def\tensor[#1]{\overset{\leftrightharpoons}{#1}}

\def\psd[#1]{\mathcal{P}_+(#1)}

\def\RR{{\mathbb R}}
\def\CC{{\mathbb C}}
\def\NN{{\mathbb N}}

\def\T[#1]{T_{#1}}

\def\Xk[#1]{{\X_{#1}}}
\def\Yk[#1]{{\Y_{#1}}}

\def\ijIdx[#1]{{{#1}_{ij}}}
\def\ikIdx[#1]{{{#1}_{ik}}}

\def\initial[#1]{{#1}^{(0)}}
\def\current[#1]{{#1}^{(n)}}
\def\next[#1]{{#1}^{(n+1)}}

\newlength\figureheight
\newlength\figurewidth
\def\fixInter{\setC}
\def\apsmInter{\Omega}
\def\modified[#1]{{\color{blue} #1}}

\newacronym{admm}{ADMM}{alternating direction method of multipliers}
\newacronym{pocs}{POCS}{projections onto convex sets}
\newacronym{apsm}{APSM}{adaptive projected subgradient method}
\newacronym{amp}{AMP}{approximate message passing}
\newacronym{oamp}{OAMP}{orthogonal approximate message passing}
\newacronym{mimo}{MIMO}{multiple-input multiple-output}
\newacronym{ser}{SER}{symbol error ratio}
\newacronym{iolama}{IO-LAMA}{individually-optimal large-MIMO AMP}
\newacronym{lmmse}{LMMSE}{minimum mean square error}
\newacronym{snr}{SNR}{signal-to-noise ratio}

\title{Superiorized Adaptive Projected Subgradient Method with Application to MIMO Detection}
\author{Jochen~Fink,~%\IEEEmembership{Student~Member,~IEEE,}
Renato~L. G.~Cavalcante,~\IEEEmembership{Member,~IEEE,}
and~S\l{}awomir Sta\'nczak,~\IEEEmembership{Senior~Member,~IEEE}% <-this % stops a space
\\
{\textit{Technische Universit\"at Berlin} and \textit{Fraunhofer Heinrich-Hertz-Institute}, Berlin, Germany}% <-this % stops a space
\\
{\{jochen.fink, renato.cavalcante, slawomir.stanczak\}@hhi.fraunhofer.de}% <-this % stops a space
\thanks{The authors acknowledge the financial support by the Federal Ministry of Education and Research of Germany in the programme of “Souverän. Digital. Vernetzt.” Joint project 6G-RIC, project identification number: 16KISK020K. The authors alone are responsible for the content of the paper.}}

\newcommand\copyrighttext{%
  \footnotesize 
  This manuscript has been submitted to IEEE Transactions on Signal Processing for possible publication.
}
\newcommand\copyrightnotice{%
\begin{tikzpicture}[remember picture,overlay]
\node[anchor=south,yshift=10pt] at (current page.south) {\fbox{\parbox{\dimexpr\textwidth-\fboxsep-\fboxrule\relax}{\copyrighttext}}};
\end{tikzpicture}%
}

\begin{document}

\maketitle\thispagestyle{empty}
\copyrightnotice{}
\begin{abstract}
In this paper, we show that the adaptive projected subgradient method (APSM) is bounded perturbation resilient. 
To illustrate a potential application of this result, we propose a set-theoretic framework for MIMO detection, and we devise algorithms based on a superiorized APSM.
Various low-complexity MIMO detection algorithms achieve excellent performance on i.i.d. Gaussian channels, but they typically incur high performance loss if realistic channel models (e.g., correlated channels) are considered.
Compared to existing low-complexity iterative detectors such as individually optimal large-MIMO approximate message passing (IO-LAMA), the proposed algorithms can achieve considerably lower symbol error ratios over correlated channels. At the same time, the proposed methods do not require matrix inverses, and their complexity is similar to IO-LAMA.
\end{abstract}
\section{Introduction}
\IEEEPARstart{S}{et-theoretic} estimation is at the heart of a large variety of signal processing techniques. 
It works by expressing any available information about the sought solution in the form of constraint sets, and by finding a feasible point, i.e., a point that is consistent with each of these constraints \cite{combettes1993foundations}. In many cases, this point can be computed with very simple algorithms based on projection methods.
A famous example is the widely used \gls{pocs} algorithm \cite{bregman1965finding,gurin1967method,stark1998vector}, which finds a point in the intersection of a finite family of closed convex sets by computing projections onto each of the sets in a cyclic manner.
The appeal of projection methods like \gls{pocs} lies in their simple structure and their potential to tackle very large problems \cite{bauschke2015projection}, \cite{censor2012effectiveness}.

To bridge the gap between this feasibility seeking approach and constrained minimization, the authors of \cite{censor2010perturbation} have proposed the superiorization methodology.
Instead of pursuing a feasible point with optimal objective value, superiorization aims at finding a feasible point with reduced objective value compared to the output of some feasibility-seeking algorithm \cite{censor2015weak}. By interleaving fixed point iterations (with respect to the feasible set) with perturbations that aim at decreasing the objective value, superiorization can reduce the computational complexity considerably compared to iterative techniques for constrained minimization \cite{censor2014projected}.
Provided that a feasibility-seeking algorithm is bounded perturbation resilient,
the convergence of its superiorized versions is guaranteed automatically \cite{censor2010perturbation}.
In recent years, bounded perturbation resilience has been proven for several classes of algorithms, including block-iterative projection methods \cite{davidi2009perturbation}, amalgamated projection methods \cite{butnariu2007stable}, dynamic string averaging projection methods \cite{censor2013convergence,bargetz2018convergence}, and fixed point iterations of averaged nonexpansive mappings \cite{he2017perturbation} (which include \gls{pocs} with relaxed projections as a particular case \cite{fink2021multi}).

The results in the above studies are restricted to problems with a finite number of constraint sets (most of them in finite dimensional spaces).
However, there exist estimation problems -- such as adaptive filtering or online learning \cite{theodoridis2010adaptive} -- in which information on the sought point arrives sequentially, resulting in feasibility problems with a potentially infinite number of constraints.
To tackle problems of this type, the authors of \cite{yamada2005adaptive} have proposed the \gls{apsm}.
This algorithmic framework is an extension of Polyak's subgradient algorithm \cite{polyak1969minimization} to the case where the cost function changes at each iteration.
Applications of the \gls{apsm} to adaptive filtering and online learning problems include multiaccess interference suppression \cite{cavalcante2008multiaccess},
acoustic feedback cancellation \cite{yukawa2006adaptive,wang2010acoustic}, robust beamforming \cite{slavakis2009adaptive}, robust subspace tracking \cite{chouvardas2014adaptive}, online radio-map reconstruction \cite{kasparick2015kernel}, kernel-based online classification \cite{slavakis2008online}, distributed learning in diffusion networks \cite{cavalcante2009adaptive,chouvardas2011adaptive,shin2018distributed}, decoding of analog fountain codes \cite{cavalcante2018low}, and adaptive symbol detection \cite{awan2018detection,awan2020adaptive}, to cite only a few.

The first objective of this study is to investigate the bounded perturbation resilience of the \gls{apsm}. We show that most of its theoretical guarantees still apply when bounded perturbations are added to the iterates in each iteration. As a result, the \gls{apsm} can be used as a basic algorithm for superiorization. 
The second objective of this study is to illustrate the usefulness of the theoretical results by applying a particular instance of a superiorized \gls{apsm} to detection in \gls{mimo} systems. In doing so, we extend our previous results \cite{fink2022set} by theoretical convergence guarantees. Nevertheless, the theoretical results on bounded perturbation resilience of the \gls{apsm} can be used in a much wider range of applications, including problems in infinite dimensional Hilbert spaces.

\subsection{Relation to Existing Studies on MIMO Detection}
\gls{mimo} detection has been studied for decades. Yet the growing interest in large-scale multi-antenna systems still drives the need for low-complexity approximation techniques.
A comprehensive overview of \gls{mimo} detection algorithms can be found in \cite{yang2015fifty} and \cite{albreem2019massive}.
The authors of \cite{jeon2015optimality} propose a low-complextity \gls{mimo} detector based on \gls{amp}. They show that this \gls{iolama} algorithm is optimal for \gls{mimo} detection over i.i.d. Gaussian channels in the large-system limit under some additional conditions. 
In \cite{ma2017orthogonal}, the authors relax the assumption of i.i.d. Gaussian channels by proposing an \gls{oamp} algorithm for \gls{mimo} detection over the more general class of unitarily invariant channel matrices. 
In contrast to the \gls{amp} detector (\gls{iolama}) proposed in \cite{jeon2015optimality}, each iteration of \gls{oamp} involves a matrix inversion in order to compute the linear \gls{lmmse} estimate, making \gls{oamp} more computationally complex than \gls{iolama}.
Many recent publications on \gls{mimo} detection \cite{samuel2017deep,samuel2019learning,gao2018sparsely,un2019deep,nguyen2020deep,khani2020adaptive,he2018model} propose deep-unfolded versions of iterative detectors.
Despite their celebrated success, some of these techniques have been found to suffer considerable performance loss on realistic channels. The authors of \cite{khani2020adaptive} mitigate this problem by proposing an online training scheme, which in turn increases the computational cost compared to deep-unfolded algorithms that are trained offline.
% As the hyperparameters used during training can influence severely the performance of the resulting deep-unfolded detectors, we restrict our attention to untrained detectors in this paper.
Although the number of iterations required to achieve good approximations can be reduced significantly by learning the algorithm parameters via deep unfolding, an extensive comparison of deep-unfolded detectors is outside the scope of this paper.
Therefore, we restrict our attention to untrained detectors.
However, we note that, owing to their iterative structure, the proposed algorithms can be readily used as a basis for deep-unfolded detection algorithms.

This study approaches \gls{mimo} detection from a set-theoretic perspective.
By posing the problem in a real Hilbert space, we devise iterative \gls{mimo} detectors with provable convergence properties based on a superiorized \gls{apsm}.
The proposed detectors have a per-iteration complexity similar to \gls{iolama}.
However, unlike \gls{iolama}, the proposed methods do not impose any assumptions on the channel matrices. 
Simulations show that, despite their low complexity, the proposed methods can outperform the more complex \gls{oamp} detector on realistic channel models specified in \cite{3gpp_tr36873}.

\subsection{Preliminaries and Notation}\label{sec:notation}
Unless specified otherwise, lowercase letters denote scalars, lowercase letters in bold typeface denote vectors, and uppercase letters in bold typeface denote matrices. The sets of nonnegative integers, nonnegative real numbers, real numbers, and complex numbers are denoted by $\NN$, $\RR_+$, $\RR$, and $\CC$, respectively.
The set of summable sequences in $\RR_+$ is denoted by $\lplusone$.
The nonnegative part of a real number $x\in\RR$ is denoted by $\relu[x]\triangleq\max\{x,0\}$. 
We denote by $\Id$ the identity operator and by $\I$ the identity matrix.
The all-zero vector is denoted by $\Null$, where the dimension of the space will be clear from the context.
Given two sets $\setA$ and $\setB$, we write $\setA\subset\setB$ or $\setB\supset\setA$ if $(\forall\x\in\setA)$ $\x\in\setB$.

Throughout this paper, we denote by $\left(\setH,\langle\cdot,\cdot\rangle\right)$ a real Hilbert space with induced norm $(\forall \x\in\setH)$ $\|\x\|:=\sqrt{\langle\x,\x\rangle}$.
Given a function $f:\setH\to\RR$, we denote by $\argmin_{\x\in\setH}f(\x)$ the set of all minimizers of $f$ (note that this set can be empty).
The distance between two points $\x,\y\in\setH$ is $d(\x,\y)=\|\x-\y\|$. The distance between a point $\x\in\setH$ and a nonempty set $\setC\subset\setH$ is defined as $d(\x,\setC)=\inf_{\y\in\setC}\|\x-\y\|$.
Following \cite{bauschke2002phase}, we define the projection of a point $\x\in\setH$ onto a nonempty subset $\setC\subset\setH$ as the set
\begin{equation*}
\Pi_\setC(\x) := \left\{\y\in\setC|~ d(\x,\y) = d(\x,\setC)\right\},
\end{equation*}
and we denote by $P_\setC:\setH\to\setC$ an arbitrary but fixed selection of $\Pi_\setC$, i.e., $(\forall \x\in\setH)$ $P_\setC(\x)\in\Pi_\setC(\x)$. If $\setC$ is nonempty, closed, and convex, the set $\Pi_\setC(\x)$ is a singleton for all $\x\in\setH$, so $\Pi_\setC$ has a unique selection $P_\setC$, which itself is called a projector. For closed nonconvex sets $\setC\neq\emptyset$ in finite-dimensional Hilbert spaces, $\Pi_\setC(\x)$ is nonempty for all $\x\in\setH$, but it is not in general a singleton. Nevertheless, we will refer to the selection $P_\setC$ as the projector, as the distinction from the set-valued operator $\Pi_\setC$ will always be clear.

The sublevel set of a function $f:\setH\to\RR$ at level $c\in\RR$ is denoted by $\lev_{\le c}f:=\{\x\in\setH~|~f(\x)\le c\}$.
We  say that a function $f:\setH\to\RR\cup\{-\infty,+\infty\}$ is coercive if $f(\x)\to+\infty$ whenever $\|\x\|\to+\infty$. Moreover, we say that the function $f$ is closed if all of its sublevel sets are closed.
In this work, we extend the notion of proximal mappings to proper closed (possibly nonconvex) functions.
\begin{definition}[Proximal Mapping]\label{def:prox}
Let $f:\setH\to(-\infty,+\infty]$ be a proper, closed function. The proximal mapping $\prox_f:\setH\to\setH$ associated with $f$ satisfies $(\forall \x\in\setH)$
\begin{equation*}%\label{eq:proximal_mapping}
    \prox_f(\x)\in \setP(\x):=\argmin_{\y\in\setH} \left(f(\y) + \frac{1}{2}\|\x-\y\|^2\right),
\end{equation*}
where we assume that $(\forall \x\in\setH)$ $\setP(\x)\neq\emptyset$.
\end{definition}
If $f$ is proper, lower-semicontinuous, and convex, the set $\setP(\x)$ is a singleton for all $\x\in\setH$, and its unique element is referred to as $\prox_f(\x)$. In this case, the assumption in Definition~\ref{def:prox} is always satisfied. If $f$ is nonconvex, we denote by $\prox_f(\x)$ a unique point selected deterministically from the set $\setP(\x)$. Note that $\setP(\x)$ is nonempty for all $\x\in\setH$ if $(\setH=\RR^N,\langle\cdot,\cdot\rangle)$ is a finite dimensional real Hilbert space and the function $\y\mapsto f(\y)+\frac{1}{2}\|\x-\y\|^2$ is coercive for all $\x\in\setH$ \cite[Theorem~6.4]{beck2017first}.

Given a subset $\setX\subset\setH$, a fixed point of a mapping $T:\setX\to\setH$ is a point $\x\in\setX$ satisfying $T(\x)=\x$. The set $\fix(T)=\{\x\in\setH~|~ T(\x)=\x\}$ is called the fixed point set of $T$.
Given two mappings
$T_1:\setH\supset\setD_1\to\setR_1\subset\setH$ and $T_2:\setH\supset\setD_2\to\setR_2\subset\setH$ such that $\setR_2\subset\setD_1$,
% $T_1, T_2:\setH\to\setH$,
we use the shorthand $T_1T_2:=T_1\circ T_2$ to denote their concatenation, which is defined by the composition $(\forall\x\in\setH)$ $T_1T_2(\x):=(T_1\circ T_2)(\x) = T_1\left(T_2(\x)\right)$. 

\begin{definition}\label{def:nonexpansive_mappings} \cite[Definition~4.1,4.33]{bauschke2011convex}, \cite{yamada2005adaptive}, \cite{he2017perturbation}
\cite[Definition~2.2.1]{cegielski2012iterative}
    Let $\setX\subset\setH$ be a nonempty subset of $\setH$.
	A mapping $T:\setX\to\setH$ is called 
	\begin{enumerate}
	    \item \emph{nonexpansive} if 
	    \begin{equation*}
	        (\forall \x\in\setX) (\forall\y\in\setX)\quad \|T(\x) - T(\y)\| \le \|\x-\y\|.
	    \end{equation*}
	    \item \emph{averaged nonexpansive} or \emph{$\alpha$-averaged nonexpansive} if there exist $\alpha\in(0,1)$ and a nonexpansive mapping $R: \setX\rightarrow \setH$ such that $T = (1-\alpha)\Id + \alpha R$.
	    \item \emph{firmly nonexpansive} ($1/2$-averaged nonexpansive) if $(\forall \x\in\setX)(\forall\y\in\setX)$
	\begin{equation*}
	    \|T(\x)-T(\y)\|^2\le \langle T(\x)-T(\y),\x-\y\rangle,
	\end{equation*}
	or equivalently, if $2T-\Id$ is nonexpansive.
	    \item \emph{quasi-nonexpansive}  if $\fix(T)\neq\emptyset$ and
	\begin{equation*}
	    (\forall \x\in\setX) (\forall \y\in\fix(T))\quad \|T(\x)-\y\|\le\|\x-\y\|.
	\end{equation*} 
	\item \emph{averaged quasi-nonexpansive} or \emph{$\alpha$-averaged quasi-nonexpansive} if there exist $\alpha\in(0,1)$ and a quasi-nonexpansive mapping $R: \setX\rightarrow \setH$ such that $T = (1-\alpha)\Id + \alpha R$.
	\item \emph{firmly quasi-nonexpansive} ($1/2$-averaged quasi-nonexpansive) if $\fix(T)\neq\emptyset$ and $(\forall \x\in\setX)(\forall\y\in\fix(T))$
	\begin{equation*}
	    \|T(\x)-\y\|^2 \le\|\x-\y\|^2- \|T(\x)-\x\|^2,
	\end{equation*}
	or equivalently, if $2T-\Id$ is quasi-nonexpansive \cite[Proposition~4.2]{bauschke2011convex}.
	\item \emph{$\kappa$-attracting quasi-nonexpansive} if $\fix(T)\neq\emptyset$ and $(\exists \kappa>0)(\forall\x\in\setX)(\forall \y\in\fix(T))$
	\begin{equation*}
     \|T(\x) - \y\|^2\le
     \|\x-\y\|^2 - \kappa\|\x-T(\x)\|^2.
     \end{equation*}
 \end{enumerate}
\end{definition}

\begin{fact}\label{fact:subdifferential}
% \cite{yamada2005adaptive}
\cite[Section~16.1]{bauschke2011convex}
Let $f:\setH\to\RR$ be a continuous
convex function\footnote{Note that convex functions $f:\setH\to\RR$ are not in general continuous if $\setH$ is infinite dimensional.}
and denote by $(\forall \x\in\setH)$
\begin{equation}\label{eq:subdifferential}
\partial f(\x):=\{\g\in\setH~|~ (\forall \y\in\setH)~ \langle\y-\x,\g\rangle+f(\x)\le f(\y)\}
\end{equation}
the \emph{subdifferential} (i.e., the set of all \emph{subgradients}) of $f$ at $\x$. Then $(\forall \x\in\setH)$ $\partial f(\x)\neq\emptyset$.
\end{fact}

\begin{fact}\label{fact:subgradient_projection}
Let $f:\setH\to\RR$ be a continuous convex function such that $\mathrm{lev}_{\le 0}f:=\{\x\in\setH~|~f(\x)\le0\}\neq\emptyset$ and let $\g(\x)\in\partial f(\x)$ be a subgradient of $f$ at $\x\in\setH$. Then the \emph{subgradient projector}
\begin{equation}\label{eq:sg_proj}
    T:\setH\to\setH: ~ \x\mapsto\begin{cases}
    \x - \frac{f(\x)}{\|\g(\x)\|^2}\g(\x) & \text{if $f(\x)>0$}\\
    \x & \text{if $f(\x)\le 0$}
    \end{cases}
\end{equation}
is \emph{firmly quasi-nonexpansive}, i.e., the mapping $2T-\Id$ is quasi-nonexpansive \cite[Proposition~2.3]{bauschke2001weak}. Moreover, the mapping $(1-\lambda)\Id + \lambda T$ is quasi-nonexpansive for all $\lambda\in[0,2]$. 
\end{fact}

\begin{fact}\cite[Proposition~2]{yamada2005adaptive}\label{fact:sg_proj_attracting}
Let $\setK\subset\setH$ be a nonempty closed convex set, and let $T$ be the subgradient projector in \eqref{eq:sg_proj} relative to be a continuous convex function $f$ with $\mathrm{lev}_{\le 0}f\neq\emptyset$. Then for any $\lambda \in (0,2)$, the mapping
\begin{equation*}
    \hat T_\lambda:=P_\setK\left((1-\lambda)\Id + \lambda T\right)
\end{equation*}
is $\left(1-\frac{\lambda}{2}\right)$-attracting quasi-nonexpansive with fixed point set $\fix(\hat T)=\setK\cup \mathrm{lev}_{\le 0}f$.
\end{fact}

\begin{definition}\label{def:bounded_perturbations}
A sequence $(\beta_n\y_n)_{n\in\NN}$ in $\setH$ is called a sequence of bounded perturbations if $(\beta_n)_{n\in\NN}\in\lplusone$ and $(\exists r\in\RR)$ $(\forall n \in\NN)$ $\|\y_n\|\le r$.
\end{definition}

\begin{definition} \cite[Definition~1.1]{combettes2001quasi} %\cite[Definition~5.32]{bauschke2011convex}
Let $\setS$ be a nonempty subset of $\setH$ and let $(\x_n)_{n\in\NN}$ be a sequence in $\setH$. Then $(\x_n)_{n\in\NN}$ is 
\begin{itemize}
    \item \emph{quasi-Fej\'er (monotone) of Type-I} relative to $\setS$ if $(\exists (\varepsilon_n)_{n\in\NN} \in \lplusone) (\forall \z\in\setS)(\forall n\in\NN)$
\begin{equation*}
     \|\x_{n+1}-\z\|\le \|\x_n-\z\| + \varepsilon_n.
\end{equation*}
\item \emph{quasi-Fej\'er (monotone) of Type-II} relative to $\setS$ if
$(\exists (\varepsilon_n)_{n\in\NN} \in \lplusone) (\forall \z\in\setS) (\forall n\in\NN)$
\begin{equation*}
    \|\x_{n+1}-\z\|^2\le \|\x_n-\z\|^2 + \varepsilon_n.
\end{equation*}
\item \emph{quasi-Fej\'er (monotone) of Type-III} relative to $\setS$ if $(\forall \z\in\setS) (\exists (\varepsilon_n)_{n\in\NN} \in \lplusone) (\forall n\in\NN)$
\begin{equation*}
     \|\x_{n+1}-\z\|^2\le \|\x_n-\z\|^2 + \varepsilon_n.
\end{equation*}
\end{itemize}
\end{definition}

The following known results related to quasi-Fej\'er monotone sequences will be used in Section~\ref{sec:apsm_bpr}. 
\begin{fact}\label{fact:quasi_fejer}
 \cite[Lemma~5.31]{bauschke2011convex}
	Let $\left(\alpha_n\right)_{n\in\NN}$ and $\left(\beta_n\right)_{n\in\NN}$ be sequences in $\RR_+$, and let $\left(\gamma_n\right)_{n\in\NN}$ and $\left(\delta_n\right)_{n\in\NN}$ be sequences in $\lplusone$ such that
	$$
	(\forall n\in\NN)\quad 
	\alpha_{n+1} \le \left(1+\gamma_n\right)\alpha_n - \beta_n + \delta_n.
	$$
	Then the sequence $\left(\alpha_n\right)_{n\in\NN}$ converges and $\sum_{n\in\NN} \beta_n$ converges.
\end{fact}

\begin{fact}\cite[Proposition~3.2]{combettes2001quasi}.\label{fact:relation_qf_types}
The different types of quasi-Fejér sequences relative to a set $\setS\subset\setH$ are related as follows.
\begin{itemize}
    \item Type-I $\implies$ Type-III
    \item Type-II $\implies$ Type-III
    \item if $\setS$ is bounded, then Type-I $\implies$ Type-II.
\end{itemize}
\end{fact}

\begin{fact}\cite[Proposition~3.2-3.3]{combettes2001quasi}.\label{fact:qf_bounded}
Let $(\x_n)_{n\in\NN}$ be a quasi-Fejér sequence (of Type-I, Type-II, or Type-III) relative to a nonempty set $\setS\subset\setH$. Then $(\x_n)_{n\in\NN}$ is bounded and $(\forall \z\in\setS)$ $(\|\x_n-\z\|)_{n\in\NN}$ converges.
\end{fact}

\section{Bounded Perturbation Resilience of the Adaptive Projected Subgradient Method}\label{sec:apsm_bpr}
In this section, we show that the \gls{apsm} \cite{yamada2005adaptive} is bounded perturbation resilient. As a consequence of this result, we enable the use of superiorized heuristics based on the \gls{apsm}. The proofs follow closely the structure
in \cite{yamada2005adaptive}, with the main difference that we add bounded perturbations to the recursions in that study. As in \cite{yamada2005adaptive}, Lemma~\ref{lem:convergence_of_projection} and Theorem~\ref{thm:strong_convergence} are technical results used to prove Theorem~\ref{thm:apsm_bpr}, the main contribution in this section. 
The following propositions are also used in the proof of Theorem~\ref{thm:apsm_bpr}.
In particular, Proposition~\ref{prop:qne_qf_type1} establishes a connection between quasi-nonexpansivity and quasi-Fej\'er monotonicity.
\begin{prop}\label{prop:qne_qf_type1}
Let $(T_n:\setH\to\setH)_{n\in\NN}$ be a sequence of quasi-nonexpansive mappings such that $\fixInter:=\bigcap_{n\in\NN}\fix(T_n)\neq\emptyset$, and let $(\beta_n\y_n)_{n\in\NN}$ be a sequence of bounded perturbations in $\setH$. Then the sequence $(\x_n)_{n\in\NN}$ generated by 
\begin{equation*}
   (\forall n\in\NN)\quad \x_{n+1} = T_n\left(\x_n + \beta_n\y_n\right),\quad \x_0\in\setH,
\end{equation*}
is quasi-Fej\'er of Type-I relative to $\fixInter$.

\emph{Proof.}
See Appendix\ref{apx:qne_qf_type1}
\end{prop}

For sequences of attracting quasi-nonexpansive mappings, we can derive a slightly stronger result stated below.

\begin{prop}\label{prop:kappa-attracting_quasi-fejer}
Let $\kappa>0$, let $(T_n:\setH\to\setH)_{n\in\NN}$ be a sequence of $\kappa$-attracting quasi-nonexpansive mappings such that $\fixInter:=\bigcap_{n\in\NN}\fix(T_n)\neq\emptyset$, and let $(\beta_n\y_n)_{n\in\NN}$ be a sequence of bounded perturbations in $\setH$. Then for any bounded subset $\setU\subset\fixInter$ the sequence $(\x_n)_{n\in\NN}$ generated by 
\begin{equation*}
    (\forall n\in\NN)\quad \x_{n+1} = T_n\left(\x_n + \beta_n\y_n\right),\quad \x_0\in\setH,
\end{equation*}
satisfies the following: $\left(\exists (\gamma_n)_{n\in\NN}\in\lplusone\right)$ $(\forall \z\in\setU)$
\begin{equation*}
    \|\x_{n+1}-\z\|^2 \le \|\x_n-\z\|^2 - \kappa\|\x_{n+1}-\x_n\|^2 + \gamma_n.
\end{equation*}

\emph{Proof.}
See Appendix\ref{apx:kappa-attracting_quasi-fejer}
\end{prop}

The following lemma is a generalization of \cite[Lemma~1]{yamada2005adaptive} to quasi-Fej\'er monotone sequences.
The proof follows almost line by line the proof in \cite[Lemma~1]{yamada2005adaptive}.

\begin{lemma}\label{lem:convergence_of_projection}
Suppose that a sequence $(\u_n)_{n\in\NN}$ in $\setH$ is quasi-Fej\'er monotone of Type-I relative to a closed convex set $\setC\subset\setH$. In addition, suppose that $\setC$ has a nonempty relative interior with respect to a linear variety $\setV\subset\setH$, i.e., there exist $\x_0\in\setC\cap\setV$ and $\varepsilon>0$ satisfying $\setU:=\{\x\in\setV~|~ \|\x-\x_0\|\le\varepsilon\}\subset\setC$. Then $(P_\setV(\x_n))_{n\in\NN}$ converges strongly to a point in $\setV$.
\emph{Proof:}
See Appendix\ref{apx:convergence_of_projection}
\end{lemma}

Lemma~\ref{lem:convergence_of_projection} is used to prove the following theorem, which generalizes \cite[Theorem~1]{yamada2005adaptive} to quasi-Fej\'er monotone sequences.
It provides sufficient conditions for strong convergence of quasi-Fej\'er monotone sequences.
The proof follows very closely the proof in \cite{yamada2005adaptive}.
\begin{theorem}\label{thm:strong_convergence}
Let $(\u_n)_{n\in\NN}$ be a quasi-Fej\'er sequence of Type-I relative to a closed convex set $\setC\subset\setH$, and suppose that there exist $\kappa>0$ and $(\gamma_n)_{n\in\NN}\in\lplusone$ such that $(\forall \z\in\setC)(\forall n \in \NN)$
\begin{equation}\label{eq:thm_conv_quasi_fejer}
    \kappa\|\u_n - \u_{n+1}\|^2 \le\|\u_n-\z\|^2 - \|\u_{n+1}-\z\|^2 + \gamma_n.
\end{equation}
Then $(\u_n)_{n\in\NN}$ converges strongly to a point in $\setH$ if $\setC$ has a nonempty relative interior with respect to a hyperplane $\setW\subset\setH$.

\emph{Proof:}
See Appendix\ref{apx:strong_convergence}
\end{theorem}

Finally, Theorem~\ref{thm:apsm_bpr}, which is based on \cite[Theorem~2]{yamada2005adaptive}, states the main result of this section. It shows that perturbed versions of the \gls{apsm} essentially enjoy the same convergence guarantees as their unperturbed counterpart in \cite{yamada2005adaptive}, except for monotone approximation. The proof of Theorem~\ref{thm:apsm_bpr} relies on Propositions~\ref{prop:qne_qf_type1} and \ref{prop:kappa-attracting_quasi-fejer}, and on Theorem~\ref{thm:strong_convergence}.

\begin{theorem}\label{thm:apsm_bpr}
Let $(\Theta_n:\setH\to\RR_+)_{n\in\NN}$ be a sequence of continuous convex functions, let $\setK\subset\setH$ be a nonempty closed convex set, and denote the \gls{apsm} update for the $n$th iteration by\footnote{The projection onto $\setK$ for $\Theta_n^\prime(\x)=\Null$ ensures that the perturbed \gls{apsm} generates a sequence in $\setK$ regardless of the perturbations. 
It is not part of the definition in \cite{yamada2005adaptive}, where the absence of perturbations guarantees that the sequence produced by the APSM is restricted to the set $\setK$.} $(\forall n\in\NN)$ 
\begin{align}\label{eq:apsm_mapping}
T_n&:\setH\to\setH\notag\\
    \x &\mapsto \begin{cases}
    P_\setK\left(\x - \lambda_n\frac{\Theta_n(\x)}{\|\Theta_n^\prime(\x)\|^2}\Theta_n^\prime(\x)\right) & \text{if $\Theta_n^\prime(\x)\neq \Null$}, \\
    P_\setK(\x) & \text{otherwise},
    \end{cases}
\end{align}
where $\Theta_n^\prime(\x_n)\in\partial\Theta_n(\x
_n)$ and $\lambda_n\in[0,2]$.
Moreover, let $(\beta_n\y_n)_{n\in\NN}\subset\setH$ be a sequence of bounded perturbations,
define $(\forall n \in\NN)$
    \begin{equation*}
         \apsmInter_n:=\left\{\x\in\setK ~\left|~ \Theta_n(\x)=\Theta_n^\star:=\underset{\x\in\setK}{\inf}~ \Theta_n(\x)\right.\right\},
    \end{equation*}
    and suppose that
     \begin{equation}\label{eq:conditions1}
    (\forall n\in\NN)\quad\Theta_n^\star=0\quad \text{and}\quad \apsmInter:={\bigcap}_{n\in\NN}\apsmInter_n\neq\emptyset.
    \end{equation}
Then for any $\x_0\in\setK$, the sequence $(\x_n)_{n\in\NN}$ in $\setK$ generated by the perturbed \gls{apsm}\footnote{We can assume without loss of generality that $\x_0\in\setK$, since $(\forall \x\in\setH)$ $T_0(\x)\in\setK$.}
\begin{equation}\label{eq:perturbed_apsm}
    \x_0\in\setK,\quad \x_{n+1} = T_n\left(\x_n + \beta_n\y_n\right)
\end{equation}
satisfies the following:
\begin{enumerate}[(a)]
   \item The sequence $(\x_n)_{n\in\NN}$ is quasi-Fejér monotone of Type-I relative to $\apsmInter$, so $(\x_n)_{n\in\NN}$ is bounded. 
    \item Moreover, if in addition to \eqref{eq:conditions1} $\left(\exists (\varepsilon_1,\varepsilon_2)\in \RR_+^2\right)$ $(\forall n\in\NN)$ $\lambda_n\in[\varepsilon_1,2-\varepsilon_2]\subset(0,2)$ and $\left(\Theta_n^\prime(\x_n+\beta_n\y_n)\right)_{n\in\NN}$ is bounded, then $\lim_{n\to\infty}\Theta_n(\x_n + \beta_n\y_n)=0$.
    \item Assume that $\apsmInter$ has a relative interior w.r.t. a hyperplane $\setW\subset\setH$, i.e., $(\exists \tilde \u\in\apsmInter\cap\setW)$ and $(\exists \varepsilon>0)$ satisfying $\setU:=\{\u\in\setW~|~ \|\u-\tilde \u\|\le \varepsilon\}\subset\apsmInter$. Then by using $(\forall n\in\NN)$ $\lambda_n\in[\varepsilon_1,2-\varepsilon_2]\subset(0,2)$, the sequence $(\x_n)_{n\in\NN}$ in \eqref{eq:perturbed_apsm} converges strongly to a point $\hat \u\in\setK$, i.e., $\lim_{n\to\infty}\|\x_n - \hat \u\|=0$.
    Moreover, $\lim_{n\to\infty} \Theta_n(\hat\u)=0$ provided that (i) $(\Theta_n^\prime(\x_n+\beta_n\y_n))_{n\in\NN}$ is bounded
    and that (ii) there exists bounded $(\Theta_n^\prime(\hat \u))_{n\in\NN}$, where $(\forall n\in\NN)$ $\Theta_n^\prime(\hat \u)\in\partial\Theta_n(\hat \u)$.
    \item In addition the assumptions (i) and (ii) in (c), assume that $\apsmInter$ has an interior point $\tilde\u$, i.e., $(\exists \rho>0)$ satisfying $\{\v\in\setH~|~ \|\v-\tilde\u\|\le\rho\}\subset\apsmInter$. Define $(\x_n)_{n\in\NN}$ by using $(\forall n\in\NN)$ $\lambda_n\in[\varepsilon_1,2-\varepsilon_2]\subset(0,2)$, and let $\hat\u:=\lim_{n\to\infty} \x_n\subset\setK$ (the existence of $\hat\u$ is guaranteed by (c)). In this case, if 
    \begin{equation*}
        (\forall\varepsilon>0)(\forall r>0)(\exists \delta>0)\quad
        \underset{\substack{d(\x_n, \lev_{\le0}\Theta_n)\ge\varepsilon \\ \|\tilde\u-\x_n\|\le r}}{\inf} \Theta_n(\x_n)\ge\delta,
    \end{equation*}
    the limit $\hat\u$ satisfies $\hat\u\in\overline{\underset{n \to \infty}{\liminf}~\apsmInter_n}$, where $\underset{n \to \infty}{\liminf}~\apsmInter_n:=\bigcup_{n=0}^\infty\bigcap_{k\ge n} \apsmInter_k$ and the overbar denotes the closure of a set.
     \end{enumerate}
     
 \emph{Proof:}
 See Appendix\ref{apx:apsm_bpr}
\end{theorem}
\begin{remark}\label{rem:initial_iterations}
We note that in \cite{yamada2005adaptive}, the condition in \eqref{eq:conditions1} does not concern the initial $n_0$ iterations, allowing for a finite number of cost functions that lead to an empty intersection of zero level sets.
Nevertheless, Theorem~\ref{thm:apsm_bpr} still covers this case if we let $\x_0:=\tilde\x_{n_0}$, where $\tilde\x_{n_0}\in\setH$ denotes the estimate after the first $n_0$ iterations. 
\end{remark}
Owing to the versatile applicability of the \gls{apsm},
Theorem~\ref{thm:apsm_bpr} can be used to prove the convergence of algorithms in a wide range of applications. For example, it offers a straightforward means of proving the convergence of the heuristic proposed in \cite{cavalcante2018low}. 
An application to channel estimation for hybrid beamforming architectures in an online setting can be found in \cite{fink2022thesis}.
Moreover, unlike many existing results on bounded perturbation resilience, Theorem~\ref{thm:apsm_bpr} even applies to infinite dimensional Hilbert spaces.
In the remainder of this paper, we use Theorem~\ref{thm:apsm_bpr} to devise iterative \gls{mimo} detectors based on a superiorized \gls{apsm}.

%%% Problem Statement %%%%%%%%%%
\section{Application to MIMO Detection}
To illustrate the usefulness of the theoretical results in Section~\ref{sec:apsm_bpr}, we devise iterative \gls{mimo} detectors with low complexity based on a superiorized \gls{apsm}. 
We consider a \gls{mimo} system with $\Nt$ transmit- and $\Nr$ receive antennas.
For square constellations (QPSK, QAM; see, e.g., \cite[Section~1.3]{wang2004wireless}), which are commonly used in practice, we can describe the system using the real-valued signal model \cite{telatar1999capacity}
\begin{equation*}
   \y = \H\s + \w, 
\end{equation*}
where $\y\in\RR^{2\Nr}$ is the received signal, $\H\in\RR^{2\Nr\times 2\Nt}$ is the channel matrix, $\s\in\RR^{2\Nt}$ is the transmit signal
with coefficients $(\forall k\in\setI:=\{1,\dots,2\Nt\})$ $s_k\in\setA\subset\RR$ drawn independently from a uniform distribution over a finite set $\setA$ of real-valued constellation points, and $\w\sim\setN(\Null,\frac{\sigma^2}{2}\I)$ is a $2\Nr$-dimensional real vector of i.i.d. Gaussian noise samples.

The goal of \gls{mimo} detection is to estimate the transmit signal vector $\s$ based on knowledge of the channel $\H$ and the received signal vector $\y$. Since the entries of $\s$ are distributed uniformly over the constellation alphabet and $\w$ is a vector of Gaussian noise, the optimal detector uses the maximum likelihood criterion given by
\begin{equation}\label{eq:ml_problem}
    \xml\in \underset{\x\in\setS}{\argmax}~ p\left(\y|\x\right) = \underset{\x\in\setS}{\argmin}~ \|\H\x - \y\|_2^2,
\end{equation}
where $\setS:=\setA^{2\Nt}\subset\RR^{2\Nt}$ is the discrete set of feasible transmit signal vectors and $p(\y|\x)$ denotes the conditional probability of $\y$ given $\x$.
The maximum likelihood problem is known to be NP-hard \cite{micciancio2001hardness} (and, in fact, NP-complete \cite{del2017mixed}). Therefore, various suboptimal approximations have been proposed.

In Section~\ref{sec:set_theoretic_formulation}, we formulate Problem~\eqref{eq:ml_problem} in a real Hilbert space, which allows us to propose algorithms with convergence guarantees based on Theorem~\ref{thm:apsm_bpr}.
In Section~\ref{sec:md:basic_alg}, we replace the finite set $\setS$ in Problem~\eqref{eq:regularized_ml} by its convex hull and we propose an \gls{apsm} to approximate a solution to the relaxed problem. Subsequently, in Section~\ref{sec:md:superiorization}, we propose superiorized version of this algorithm by adding bounded perturbations in each iteration with the intent to steer the iterate towards a solution to the nonconvex maximum likelihood problem. 
Similarly to \gls{amp}, which alternates between gradient steps and (Gaussian) denoising steps, the proposed algorithm interleaves subgradient projections onto sublevel sets with denoising steps defined by hard slicing or soft thresholding. 
A convergence proof for the proposed method is provided in Section~\ref{sec:md:convergence}, and the algorithmic steps are summarized in Section~\ref{sec:alg_summary}.

%%% Algorithmic Solution %%%%%%%%%%%%%%%%%%%%%%%%%%%%%%%%%%%%%%%%%%%%%%%%%%%%%%%%%%%%%%%
\subsection{Set-theoretic Formulation in a Hilbert Space}\label{sec:set_theoretic_formulation}
In the following, we approach the problem from a set-theoretic perspective, which enables us to devise low-complexity approximation techniques with provable convergence properties without imposing any additional assumptions.
To apply the results in Section~\ref{sec:apsm_bpr}, we formulate Problem~\eqref{eq:ml_problem} in a real Hilbert space $\left(\setH:=\RR^{2\Nt}, \langle\cdot,\cdot\rangle\right)$ equipped with the standard Euclidean inner product
\begin{equation*}
    (\forall \x,\y\in\setH)\quad \langle\x,\y\rangle:=\y^T\x,
\end{equation*}
which induces the Euclidean norm
% $\|\x\|=\sqrt{\langle\x,\x\rangle}$.
% $\|\cdot\|=\sqrt{\langle\cdot,\cdot\rangle}=\|\cdot\|_2$.
$\|\cdot\|=\|\cdot\|_2$.
In this Hilbert space, we can express the maximum likelihood problem in \eqref{eq:ml_problem} as 
\begin{equation}
    \label{eq:regularized_ml}
    \underset{\x\in\setH}{\minimize}~ \|\H\x-\y\|^2 + \iota_{\setS}(\x),
\end{equation}
where and $\iota_{\setS}:\setH\to\RR_+\cup\{+\infty\}$ is the indicator function of $\setS$ given by
\begin{equation*}
    (\forall \x\in\setH)\quad \iota_{\setS}(\x)=\begin{cases}
    0 & \text{if $\x\in\setS$}\\
    +\infty & \text{otherwise}.
    \end{cases}
\end{equation*}

\subsection{An Adaptive Projected Subgradient Method for MIMO Detection}\label{sec:md:basic_alg}
In principle, a solution to the maximum likelihood problem can be approximated with iterative techniques that interleave gradient steps for the cost function with projections onto the nonconvex constraint set in \eqref{eq:ml_problem}. Such algorithms, based on projected gradient methods or the \gls{admm}, have been discussed in \cite{liu2017discrete, samuel2019learning,un2019deep}. However, owing to the projection onto the nonconvex constellation alphabet, convergence of these algorithms cannot be guaranteed without imposing stringent assumptions on the channel matrix (see \cite{liu2017discrete}).
Instead of directly approaching the nonconvex maximum likelihood problem in \eqref{eq:ml_problem}, some authors \cite{tan2001constrained, thrampoulidis2016ber, wu2016high} have applied iterative algorithms to a relaxed version of Problem \eqref{eq:ml_problem}, in which the discrete set $\setS$ is replaced with its convex hull 
\begin{equation*}
    \setB:=\left\{\x\in\setH~|~\|\x\|_\infty\le a_{\max}\right\},
\end{equation*}
where $a_{\max} = \underset{a\in\setA}{\max}~ |a|$.
The resulting suboptimal detector
\begin{equation}\label{eq:box_problem}
\hat\s\in\underset{\x\in\setB}{\argmin}~ \|\H\x-\y\|^2
\end{equation}
is also referred to as box-relaxation decoder \cite{thrampoulidis2018symbol}.
In the following, we devise a basic algorithm based on an \gls{apsm} that aims at approximating solutions to Problem~\eqref{eq:box_problem}.
% minimizing the objective of \eqref{eq:ml_problem} over the closed convex set $\setB$. 
According to Theorem~\ref{thm:apsm_bpr} and Remark~\ref{rem:initial_iterations}, convergence of the \gls{apsm} can only be guaranteed if all but finitely many of its cost functions attain the value zero. Hence we cannot directly use the objective function in \eqref{eq:ml_problem} as a cost function for all iterations of the \gls{apsm}.
If the optimal objective value $\rhohat:=\|\H\hat\x-\y\|^2$ of Problem~\eqref{eq:box_problem} were known, we could use the \gls{apsm} to solve the problem
\begin{equation*}
    \underset{\x\in\setB}{\minimize}~\left(\|\H\x-\y\|^2-\rhohat\right)_+.
\end{equation*}
This reformulation of the convex minimization problem in \eqref{eq:box_problem} is equivalent to the convex feasibility problem
\begin{equation}\label{eq:cfp}
    \find~\x ~\text{such that}~ \x\in\setC_{\rhohat}\cap\setB,
\end{equation}
where $(\forall\rho\ge 0)$
$\setC_\rho:=\left\{\x\in\setH~|~ \|\H\x - \y\|^2 \le \rho  \right\}$
is a sublevel set of the objective function in \eqref{eq:ml_problem}, also known as stochastic property set \cite{yamada2002efficient}.
In the following, we build upon a technique shown in \cite{cavalcante2018low}, where the objective is to solve the problem
\begin{equation}\label{eq:cfp2}
    \find~\x ~\text{such that}~
    \x\in\left(\bigcap_{n\ge n_0}\setC_{\rho_n}\right)\cap\setB,
\end{equation}
for some $n_0\in\NN$, given a sequence $\left(\setC_{\rho_n}\right)_{n\in\NN}$ of stochastic property sets. 
As in \cite{cavalcante2018low}, we define a sequence of continuous convex functions $\Theta_n: \setH\to\RR_+$ by
\begin{equation*}
    (\forall n\in\NN)(\forall x\in\setH)\quad 
    \Theta_n(\x) := \left(\|\H\x-\y\|^2-\rho_n\right)_+
\end{equation*}
and we use the \gls{apsm} to minimize asymptotically (in the sense defined in \cite[Theorem~2(b)]{yamada2005adaptive}) this sequence of functions over the set $\setB$ by iteratively applying the recursion
\begin{equation}\label{eq:apsm_update_mimo}
    \x_0\in\setH,\quad (\forall n\in\NN)~ \x_{n+1}:=T_n(\x_n),
\end{equation}
where
\begin{equation}\label{eq:md:apsm_mapping}
    T_n(\x) :=\begin{cases}
    P_\setB\left(\x-\mu_n\frac{\Theta_n(\x)}{\left\|\Theta_n^\prime(\x)\right\|^2}\Theta_n^\prime(\x)\right) & \text{if $\Theta_n(\x)>0$}\\
    P_\setB(\x) & \text{otherwise.}
    \end{cases}
\end{equation}
Here, $(\forall n\in\NN)$ $\Theta_n^\prime:\setH\to\setH: \x\mapsto 2\H^T(\H\x-\y)\in\partial\Theta_n(\x)$ defines a subgradient of $\Theta_n$ at $\x$, and $\mu_n\in[\varepsilon_1,2-\varepsilon_2]\subset(0,2)$ is a relaxation parameter.
If we choose the elements of $(\rho_n)_{n\in\NN}$ to increase monotonically in such a way
that $(\exists n_0\in\NN)$ $\rho_{n_0}>\rhohat$,
% $\lim_{n\to\infty}~\min_{\x\in\setH}~\|\H\x-\y\|^2-\rho_n\le 0$, 
the recursion in \eqref{eq:apsm_update_mimo} is guaranteed to converge (see Section~\ref{sec:md:convergence}). 
Moreover, if $\rho_0$ is sufficiently small and $(\rho_n)_{n\in\NN}$ increases sufficiently slowly, the final objective value $\lim_{n\to\infty}\|\H\x_n-\y\|^2$ will be close to optimal. In the next subsection, we devise superiorized versions of the algorithm in \eqref{eq:apsm_update_mimo}, which additionally aim at enforcing the nonconvex constraint $\setS$. As the optimal objective value $\rhoML:=\|\H\xml-\y\|^2$ of the ML problem in \eqref{eq:ml_problem} may be larger than $\rhohat$, we additionally require the sequence $(\rho_n)_{n\in\NN}$ to satisfy $(\exists n_0\in\NN)(\forall n\ge n_0)$ $\rho_n\ge\rhoML\ge\rhohat$ in the following.
\begin{comment}
    Problem~\eqref{eq:cfp2} is feasible only for $n_0$ such that $(\forall n\ge n_0)$ $\rho_n\ge\rhohat$. Moreover, for the feasible set to contain the ML estimate $\xml$, it is required that $(\forall n\ge n_0)$ $\rho_n\ge\rhoML$. Furthermore, the algorithm in \eqref{eq:apsm_update_mimo} will make progress (in the sense that $\x_{n+1}\neq\x_n$) only if $\rho_n < \|\H\x_n-\y\|^2$. Therefore, the elements of the sequence $(\rho_n)_{n\in\NN}$ should increase slowly, in order to ensure that there are sufficiently many iterations between the first time $\rho_n\ge\rhohat$ (or $\rho_n\ge\rhoML$, respectively) and the instant at which $\rho_n\ge\|\H\x_n-\y\|^2$.
\end{comment}

\subsection{Superiorization}\label{sec:md:superiorization}
Replacing the discrete constellation alphabet $\setS$ with its convex hull $\setB\supset \setS$ can potentially limit the performance of the algorithm in \eqref{eq:apsm_update_mimo}, because it ignores available information on the prior distribution of $\x$. 
Therefore, we use the \gls{apsm} in \eqref{eq:apsm_update_mimo} as a \emph{basic algorithm} for superiorization, and we devise a \emph{superiorized version} 
\begin{equation}\label{eq:sup_alg}
    (\forall n\in\NN)\quad \x_{n+1}:=T_n(\x_n + \beta_n\v_n),\quad \x_0\in\setH
\end{equation}
of this algorithm by adding small perturbations to its iterates with the intent to reduce slightly the value of a certain superiorization objective.
According to Theorem~\ref{thm:apsm_bpr}, convergence of the sequence generated by the recursion in \eqref{eq:sup_alg} can still be guaranteed, given that $(\beta_n\v_n)_{n\in\NN}$ are bounded perturbations in the sense of Definition~\ref{def:bounded_perturbations}, i.e., that $(\beta_n)_{n\in\NN}\in\lplusone$ and that $(\v_n)_{n\in\NN}$ is a bounded sequence in $\setH$. 
Potential choices for the sequence $(\v_n)_{n\in\NN}$ are introduced below.

Objective functions for superiorization are typically convex. Nevertheless, we consider nonconvex objective functions in the following. Moreover, as in \cite{fink2021multi}, we slightly deviate from \cite{censor2015weak} and \cite{censor2010perturbation}, by using proximal mappings instead of subgradients of the superiorization objective to define the perturbations.
In this way, we enable a simple trade-off between the perturbations' magnitude and their contribution to reducing the objective value.
To incorporate prior information about the transmit signal, we are interested in superiorization objective functions $f:\setH\to\RR_+\cup\{+\infty\}$ that satisfy $f(\x)=0$ if and only if $\x\in\setS$.
One example of such a function is the indicator function $\fltwo:=\iota_\setS$.
The proximal mapping associated with $\fltwo$ is given by
$\prox_{\fltwo}(\x) = P_{\setS}(\x)$.
Here, $P_{\setS}$ denotes a projection onto the set $\setS$. Since $\setS$ is not convex, this point is not unique for all $\x\in\setH$. However, a projection onto $\setS$ always exists because the set is closed and the space $\setH$ is finite dimensional.
In this way, we can devise perturbations of the form 
\begin{equation}\label{eq:pert_l2}
    (\forall n\in\NN)\quad \vltwo:=P_{\setS}(\x_n)-\x_n.
\end{equation}

As the primary objective of \gls{mimo} detection is to reduce the \gls{ser}, one could instead use a superiorization objective that penalizes the number of coefficients of the estimate $\hat\x\in\setH$ that lie outside of the set of valid constellation points, i.e.,
\begin{equation}\label{eq:ell_0}
    \sum_{k:\hat x_k\notin\setA} 1 = \|\hat\x - P_{\setS}(\hat\x)\|_0,
\end{equation}
where $\|\cdot\|_0$ denotes the $\ell_0$ pseudo-norm. 
Borrowing a well-known technique from compressed sensing \cite{donoho2006compressed}, we replace the $\ell_0$ pseudo-norm in \eqref{eq:ell_0} with the $\ell_1$-norm to define an alternative superiorization objective
$
(\forall \x\in\setH)\quad \flone(\x):=\|\x-P_{\setS}(\x)\|_1.  
$
Note that $\flone$ is still nonconvex due to the projection onto the nonconvex set $\setS$. 
Nevertheless, based on \cite[Example~6.8]{beck2017first} and the translation property of the proximal mapping \cite[Theorem~6.11]{beck2017first},
$(\forall \tau \ge 0)$ we can define a proximal mapping associated with $\tau \flone$ by\footnote{A detailed proof can be found in \cite[Appendix~B]{fink2022thesis}.}
$
    \prox_{\tau \flone}(\x) = \shrink_\tau\left(\x - P_{\setS}(\x)\right) + P_{\setS}(\x),
$
where $(\forall \tau \ge 0)$ $\shrink_\tau:\setH\to\setH$ is the soft-thresholding operator
\begin{equation*}
    (\forall \x\in\setH)(\forall k\in \setI)~ \shrink_\tau(\x)|_k:=\sign(x_k)(|x_k|-\tau)_+.
\end{equation*}
As a result, we obtain perturbations of the form $(\forall n\in\NN)$
\begin{align}\label{eq:pert_l1}
    \vlone&:=\prox_{\tau_n \flone}(\x_n)-\x_n\notag\\
    &=\shrink_{\tau_n}\left(\x _n- P_{\setS}(\x_n)\right) + P_{\setS}(\x_n)-\x_n.
\end{align}
The convergence of the superiorized \gls{apsm} in \eqref{eq:sup_alg} with perturbations according to \eqref{eq:pert_l2} or \eqref{eq:pert_l1} is investigated below. 

\subsection{Convergence of the Proposed Algorithms}\label{sec:md:convergence}
% In the following, we investigate the convergence of the proposed algorithms. 
According to Theorem~\ref{thm:apsm_bpr} and Remark~\ref{rem:initial_iterations}, the sequence produced by the superiorized \gls{apsm} in \eqref{eq:sup_alg} converges (strongly) to a point $\x^\star\in\setB$, given that the perturbations are bounded and that
\begin{enumerate}[(C1)]
    \item $(\exists n_0\in\NN)(\forall n \ge n_0)$ $\Theta_n^\star= 0$, i.e., defining $(\forall n\in\NN)$ $\apsmInter_n:=\{\x\in\setB~|~ \Theta_n(\x)=\Theta_n^\star\}=\setB\cap\setC_{\rho_n}$, we have $\apsmInter:=\bigcap_{n\ge n_0}\apsmInter_n\neq\emptyset$.
    \item $(\exists \z\in\apsmInter)(\exists\eta>0)$ $\left\{\x\in\setH~|~ \|\x-\z\|\le\eta\right\}\subset\apsmInter$, i.e., the set $\apsmInter$ has an interior point.
\end{enumerate}
Moreover, the point $\x^\star$ minimizes all but finitely many functions of the sequence $(\Theta_n)_{n\in\NN}$ if
\begin{enumerate}[(C1)]
\setcounter{enumi}{2}
    \item the sequence $\left(\Theta_n^\prime(\x_n+\beta_n\y_n)\right)_{n\in\NN}$ is bounded
    \item there exists a bounded sequence $\left(\Theta_n^\prime(\x^\star)\right)_{n\in\NN}$, where $(\forall n\in\NN)$ $\Theta_n^\prime(\x^\star)\in\partial\Theta_n(\x^\star)$
\end{enumerate}

The objective of the remainder of this subsection is to show that these conditions are satisfied.
We begin by showing that the proposed perturbations are bounded.
\begin{prop}
The proposed perturbations
in \eqref{eq:pert_l2} and \eqref{eq:pert_l1} are bounded.

\emph{Proof:}\label{prop:pert_bounded}
Since $\setB$ is compact, we can define $c:=\max_{\x\in\setB}\|\x\|$. By \eqref{eq:md:apsm_mapping} and the definition of a projection, $(\forall n\in\NN)$ $\x_n\in\setB$ and $(\forall \x\in\setH) P_\setS(\x)\in\setS\subset\setB$. Consequently, we have
\begin{equation*}
    \|\vltwo\| = \|P_{\setS}(\x) - \x_n\|\le \|P_{\setS}(\x)\| + \|\x_n\|\le 2c
\end{equation*}
and
\begin{align*}
    \|\vlone\|&= \|\shrink_\tau\left(\x - P_{\setS}(\x)\right) + P_{\setS}(\x) - \x_n\| \notag\\
    &\le\|\shrink_\tau\left(\x - P_{\setS}(\x)\right)\|+ \|P_{\setS}(\x) - \x_n\|\notag\\
    &\le 2\|P_{\setS}(\x) - \x_n\|\le 4c, 
\end{align*}\pushQED{\qed}
which concludes the proof.\qedhere
\popQED
\end{prop}

Finally, we use Theorem~\ref{thm:apsm_bpr} and Proposition~\ref{prop:pert_bounded} to prove the convergence of the proposed algorithms.
Since the proof requires the convex hull of the constellation constraint to have an interior point, we restrict our attention to constellation alphabets $\setA$ the elements of which span the entire complex plane. Note that we can still apply the proposed approach to other constellations, such as BPSK (see, e.g., \cite[Section~1.3]{wang2004wireless}), by posing the problem in a subspace of $\setH$.
\begin{prop}
Let  $\left(\rho_n\right)_{n\in\NN}$ be a sequence in $\RR_+$ satisfying $(\exists \nmin\in\NN)$ $(\exists\eta>0)$ $(\forall n\ge \nmin)$ $\rho_{n}\ge\rhoML+\eta$.
Then the algorithm in \eqref{eq:sup_alg} with $(\beta_n)_{n\in\NN}\in\lplusone$ and perturbations
% $\left(\vltwo\right)_{n\in\NN}$ or $\left(\vlone\right)_{n\in\NN}$
according to \eqref{eq:pert_l2} or \eqref{eq:pert_l1} is guaranteed to converge to a point $\x^\star\in\setB$ minimizing all but finitely many functions of the sequence $(\Theta_n)_{n\in\NN}$.

\emph{Proof:}
In light of Theorem~\ref{thm:apsm_bpr}, it remains to show that the conditions (C1)--(C4) above are satisfied. Let $\s^\star$ denote a solution to Problem~\eqref{eq:ml_problem}. 
\begin{enumerate}[(C1)]
    \item By assumption, $(\forall n\ge\nmin)$ $\rho_n\ge\rho^\star=\|\H\s^\star-\y\|^2$, whereby $0\le\Theta_n^\star\le\left(\|\H\s^\star-\y\|^2-\rho_n\right)_+\le0$. Moreover, $(\forall n\ge\nmin)$ $\s^\star\in\setC_{\rho_n}\cap\setB=\apsmInter_n$, and thus $\apsmInter\neq\emptyset$.
     \item Define $\setE:=\{\x\in\setH~|~\|\xml-\x\|\le\varepsilon\}$ with some positive $\varepsilon\le\frac{\sqrt{\rhoML+\eta}-\sqrt{\rhoML}}{\|\H\|_2}$. All $\u\in\setH$ with $\|\u\|\le1$ satisfy
    \begin{small}
    \begin{align*}
        \|\H(\xml+\varepsilon\u)+\y\|^2 &= \rhoML + 2\varepsilon\langle\H\xml-\y,\H\u\rangle + \varepsilon^2\|\H\u\|^2\\
        &\overset{(i)}{\le} \rhoML + 2\varepsilon\sqrt{\rhoML}\|\H\|_2 + \varepsilon^2\|\H\|_2^2\\
        &\le\rhoML+\eta,
    \end{align*}
    \end{small}
    where (i) is an application of the Cauchy-Schwartz inequality.
    Therefore, by assumption, $(\forall n\ge \nmin)$ $(\forall \x\in\setE)$ $\Theta_n(\x)=0$, i.e., $\setE\subset\setC_{\rho_n}$.
    Now, we define a set with nonempty interior by $\setQ:=\{\x\in\setH~|~(\forall k\in\setI)~ s_l \le x_i \le s_u\}$, where $(\forall k\in\setI)$ 
    $$\tilde s_k := \sign(s^\star_k)\cdot\left(|s^\star_k|-\frac{\varepsilon}{\sqrt{2\Nt}}\right),$$ 
    $s_l:=\min(\tilde s_k,s_k)$, and $s_u:=\max(\tilde s_k,s_k)$. Note that $(\forall n\ge n_0)$ $\setQ\subset\setE\subset\setC_{\rho_n}$. Moreover, $\setQ\subset\setB$ for sufficiently small $\varepsilon>0$, so it holds that $\setQ\subset\apsmInter$.
    \item Let $(\forall n\in\NN)$ $\z_n:=\x_n+\beta_n\y_n$. Since $(\forall n\in\NN)$ $\Theta_n(\x)=0 \implies \Theta_n^\prime(\x) = \Null$, it is sufficient to consider the case $\Theta_n(\x)>0$. In this case, we have that $\Theta_n^\prime(\x)=2\H^T(\H\x-\y)$, so  $(\forall \x_n\in\setB)$
    \begin{align*}
        \|\Theta_n^\prime(\z_n)\| &\overset{(i)}{\le} 2\|\H^T\H\z_n\| + 2\|\H^T\y\|\\
        &\overset{(ii)}{\le} 2\|\H^T\H\|_2\cdot\|\z_n\| + 2\|\H^T\y\|\\
         &\overset{(iii)}{\le} 2\|\H^T\H\|_2\cdot\left(\|\x_n\|+\beta_n\|\y_n\|\right) + 2\|\H^T\y\|\\
        &\overset{(iv)}{\le} 2 \left(c+r\max_{n\in\NN}\beta_n\right)\|\H^T\H\|_2 + 2\|\H^T\y\|.
    \end{align*}
    Here,  (i) and (iii) follow from the triangle inequality, (ii) follows from the definition of an operator norm, and (iv) follows from the definition of the constant $c$ in Proposition~\ref{prop:pert_bounded} and the fact that $(\exists r\in\RR)$ $(\forall n\in\NN)$ $\|\y_n\|\le r$. Consequently, the sequence of subgradients $\left(\Theta_n^\prime(\x_n+\beta_n\y_n)\right)_{n\in\NN}$ is bounded.
    \item Since $(\x_n)_{n\in\NN}$ is a convergent sequence in the compact set $\setB$, its limit $\x^\star$ also belongs to $\setB$. Therefore, we can apply the same argument as above.\pushQED{\qed}\qedhere\popQED
\end{enumerate}
\end{prop}
\begin{comment}
 Note that the convergence proof in this subsection does not depend on the choice of the channel matrix $\H$. Hence, the proposed algorithms are guaranteed to converge for arbitrary channel matrices.
\end{comment}

\subsection{Summary of the Proposed Algorithms}\label{sec:alg_summary}
The proposed iterative \gls{mimo} dectectors with perturbations according to \eqref{eq:pert_l2} and \eqref{eq:pert_l1}, respectively, are summarized in Algorithm~\ref{alg:sapsm} below.
\begin{algorithm}[H]
\caption{Superiorized APSM for MIMO Detection}\label{alg:sapsm}
\begin{algorithmic}[1]
	\State \textbf{Parameters:}~ $(\rho_n\ge 0)_{n\in\NN}$, $(\mu_n\in(0,2))_{n\in\NN}$, $(\beta_n\ge 0)_{n\in\NN}$, $(\tau_n\ge 0)_{n\in\NN}$
	\State \textbf{Input:}~ $\H\in\RR^{2\Nr\times 2\Nt}, \y\in\RR^{2\Nr}$
	\State \textbf{Output:}~ $\hat\x\in\RR^{2\Nt}$
	\State \textbf{Initialization:}~ Choose arbitrary $\x_0\in\setH$ %\Comment{E.g., $\X^{(0)}\gets\Null$}
	\For{$n=0,\dots,\nmax-1$}	
	\State {\small{}$\v_n=\prox_{\tau_n f_{\ell_{j}}}(\x_n) - \x_n$, $j\in\{1,2\}$}\Comment Eq. \eqref{eq:pert_l2} or \eqref{eq:pert_l1}
	\State $\z_n=\x_n+\beta_n\v_n$ %\Comment compute perturbed estimate
	\State $\Theta_n(\z_n)=(\|\H\z_n-\y\|^2-\rho_n)_+$ %\Comment evaluate cost
	\State $\Theta_n^\prime(\z_n)=2\H^T(\H\z_n-\y)$ %\Comment compute gradient 
	\State \begin{footnotesize}$\x_{n+1} = \begin{cases} P_\setB\left(\z_n-\mu_n\frac{\Theta_n(\z_n)}{\|\Theta_n^\prime(\z_n)\|^2}\Theta_n^\prime(\z_n)\right) & \text{if $\Theta_n(\z_n)>0$}\\
	P_\setB(\z_n) & \text{otherwise}\end{cases}$ \end{footnotesize} %\Comment update
	\EndFor
	\State	\textbf{return} $\hat\x=\x_{n+1}$
\end{algorithmic}
\end{algorithm}

In the simulations in Section~\ref{sec:simulations}, we terminate the algorithm once a certain number of iterations is exceeded. Since the sequence $(\x_n)_{n\in\NN}$ is guaranteed to converge, we could alternatively terminate the algorithm once $\|\x_{n+1}-\x_n\|\le \epsilon$ for some $\epsilon>0$.
\begin{comment}
 The complexity of Algorithm\ref{alg:sapsm} is determined by the vector-matrix multiplications in lines 8 and 9, which can be computed with $\bigo(NK)$ multiplications. 
The same also holds for \gls{iolama}.
By contrast, computing the \gls{lmmse} estimate and the iterative steps of \gls{oamp} requires a matrix inverse or, for numerical stability, solving a system of linear equations using standard matrix factorization techniques. One of the best known algorithms for this step \cite{coppersmith1987matrix} has a complexity of $\bigo(K^{2.376})$.
\end{comment}

%%% Numerical Results %%%%%%%%%%%%%%%%%%%%%%%%%%%%%%%%%%%%%%%%%%%%%%%%%%%%%%%%%%%%%%%%%%%%%%%%%%%%%%%%%%

\section{Numerical Results}\label{sec:simulations}
In this section, we compare the performance of the proposed algorithms and existing methods. As a baseline, we consider the widely used \gls{lmmse} estimator given by
\begin{equation}\label{eq:lmmse}
     \x_\lmmse = (\H^T\H + \sigma^2\I)^{-1}\H^T\y.
\end{equation}
For higher-order modulations, the \gls{ser} depends on the scaling of the estimate. Therefore, the constrained version of the \gls{lmmse} estimator (also known as linearly constrained minimum variance estimator) \cite{frost1972algorithm} typically achieves lower \gls{ser} than the unconstrained \gls{lmmse} estimator in \eqref{eq:lmmse}. The constrained \gls{lmmse} estimate is given by
\begin{equation}\label{eq:constrained_lmmse}
    \tilde\x_\lmmse = \diag(\bm{\alpha})(\H^T\H + \sigma^2\I)^{-1}\H^T\y,
\end{equation}
where
\begin{equation*}
   (\forall k\in\setI) \quad \alpha_k = \left(\h_k^T\left(\H\H^T + \sigma^2\I\right)^{-1}\h_k\right)^{-1}.
\end{equation*}
Here, $\h_k$ denotes the $k$th column of $\H$.
The simulations below assess the performance of the following algorithms:
\begin{itemize}
    \item The \gls{apsm} basic algorithm in \eqref{eq:apsm_update_mimo} (\apsm)
    \item The superiorized \gls{apsm} in \eqref{eq:sup_alg} with perturbations according to \eqref{eq:pert_l2} (\apsmltwo)
    \item The superiorized \gls{apsm} in \eqref{eq:sup_alg} with perturbations according to \eqref{eq:pert_l1} (\apsmlone)
    \item The \gls{amp}-based \gls{mimo} detector (\gls{iolama}) proposed in \cite{jeon2015optimality} (\amp)
    \item The detector based on \gls{oamp} \cite{ma2017orthogonal} (\oamp)
    % \item The \gls{lmmse} estimate given by $\x_\lmmse = (\H^T\H + \sigma^2\I)^{-1}\H^T\y$ (\lmmse).
    \item The constrained \gls{lmmse} estimate $\tilde\x_\lmmse$ in \eqref{eq:constrained_lmmse}  (\lmmse)
    \item The box-relaxation decoder in \eqref{eq:box_problem}, which is computed using a general-purpose convex solver (\boxdetector).
\end{itemize}
 We consider a system with $\Nt = 16$ single antenna transmitters and $\Nr=64$ receive antennas and 16-QAM constellation. As in \cite{khani2020adaptive}, we assume perfect power allocation, i.e., we normalize the columns of the channel matrix $\H$ to unit 2-norm.
For the \gls{apsm} algorithms, we set $(\forall n\in\NN)$ $\rho_n={5\cdot 10^{-5}\cdot 1.06^n}$ and $\mu_n=0.7$. The perturbations of \apsmltwo{} are scaled using the sequence $(\beta_n=b^n)_{n\in\NN}$ with $b=0.9$. For \apsmlone{}, we set $(\forall n\in\NN)$ $\tau_n=0.005$ and $\beta_n=0.9999$. 
All iterative algorithms are initialized with $\x_0=\Null$. The design parameters of the APSM-based algorithms were chosen somewhat arbitrarily, in a way that resulted in good performance for all problem types under consideration. We note that the number of iterations required by these methods could potentially be reduced by tuning their parameters for a particular problem setting.

\begin{figure}[ht]
	\centering
	\includegraphics[width=\figurewidth]{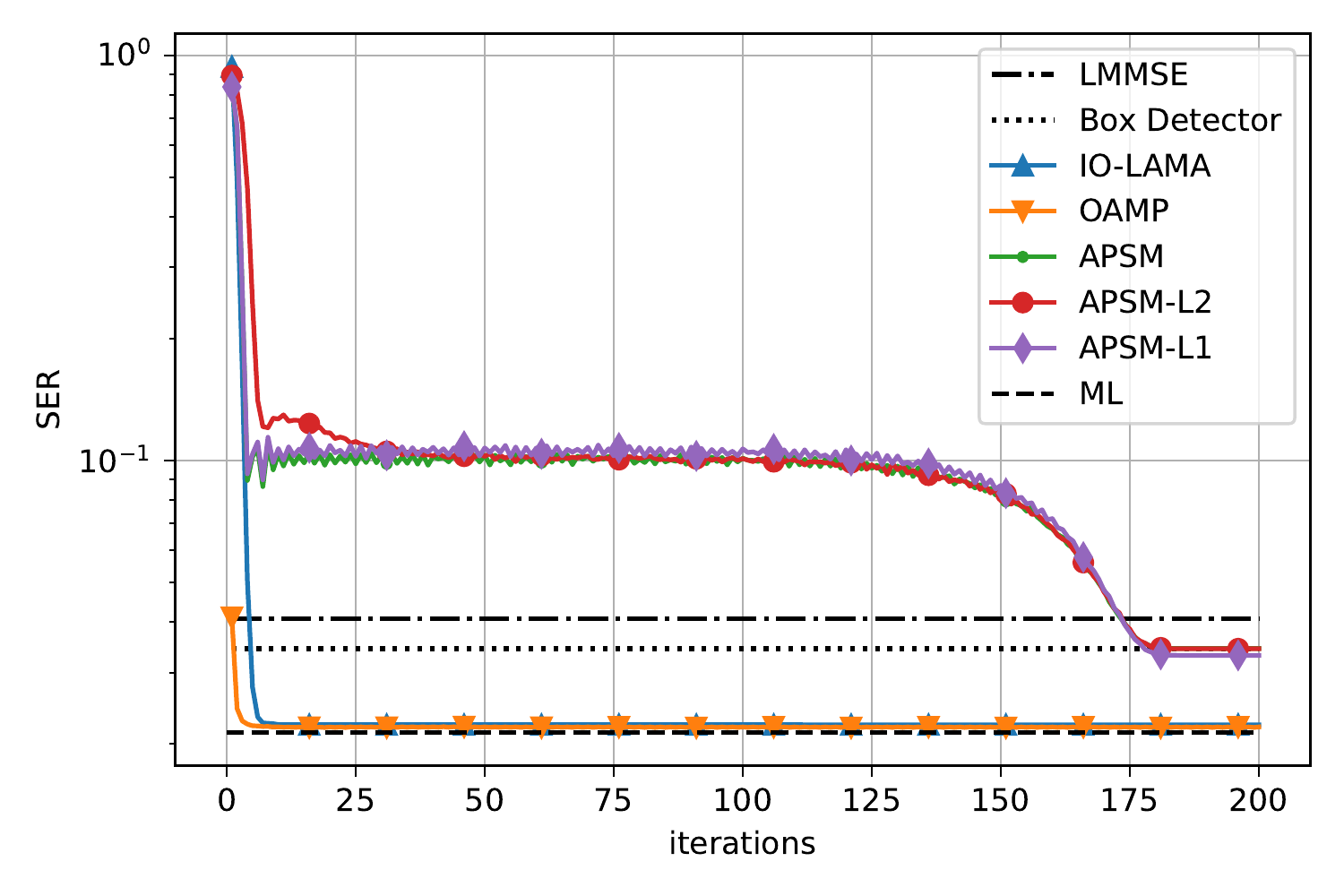}
	\caption{\gls{ser} as a function of the number of iterations averaged over 10000 realizations of i.i.d. Gaussian channels.}
	\label{fig:apsm_amp_iid}
\end{figure}

Figure~\ref{fig:apsm_amp_iid} shows the \gls{ser} throughout the iterations, averaged over 10000 i.i.d. Gaussian channel matrices with \SI{9}{dB} \gls{snr}. It can be seen that both \amp{} and \oamp{} achieve maximum likelihood performance within about 10 iterations. 
%The proposed methods do not achieve maximum likelihood performance. %However, they still outperform \lmmse{}. 
%However, both \apsm{} and \apsmltwo{} 
The proposed \apsm{} and \apsmltwo{} detectors eventually achieve the same \gls{ser} as the \boxdetector{}. The \apsmlone{} detector even achieves a slightly lower \gls{ser}.

\begin{figure}[ht]
	\centering
	\includegraphics[width=\figurewidth]{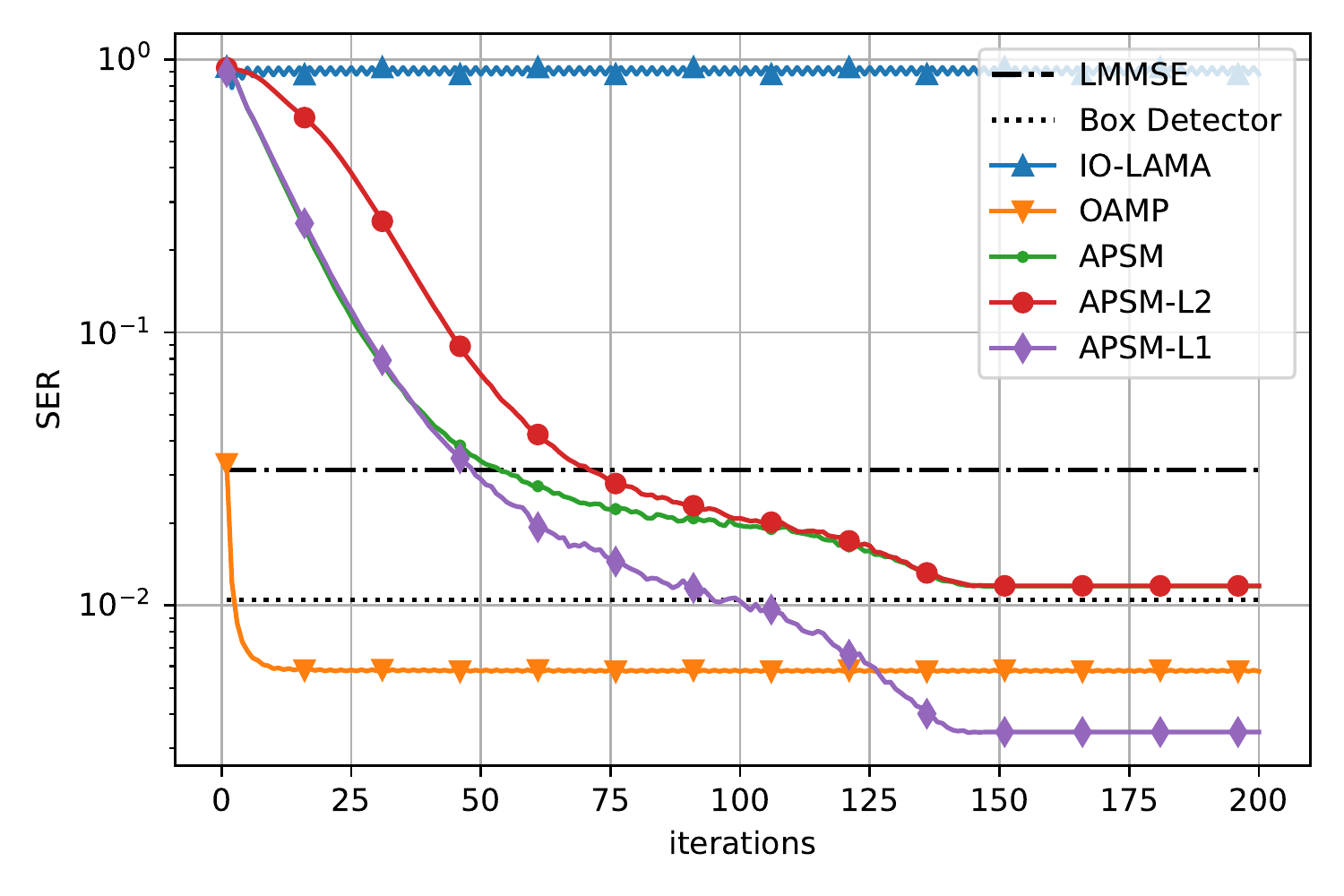}
	\caption{\gls{ser} as a function of the number of iterations averaged over 10000 3GPP channels.}
	\label{fig:apsm_amp_3gpp}
\end{figure}

Figure~\ref{fig:apsm_amp_3gpp} shows the \gls{ser} as a function of the number of iterations averaged over 10000 (single-subcarrier) 3GPP channels \cite{3gpp_tr36873} with \SI{18}{dB} \gls{snr}.
% The channels are generated with the Quasi Deterministic Radio Channel Generator (QuaDRiGa) \cite{quadriga} using the code provided with \cite{khani2020adaptive}. The simulation adopts the \texttt{3GPP\_3D\_UMa\_NLOS} scenario for a \gls{bs} equipped with a $4\times 8$ dual polarized antenna array ($\Nr=64$) according to \cite{3gpp_tr36873} and $\Nt=16$ \glspl{ue} equipped with omnidirectional antennas.
The single-subcarrier channels are drawn  at random from a dataset that was generated using the code provided with \cite{khani2020adaptive}.
While all \gls{apsm}-type algorithms achieve a \gls{ser} below \lmmse, \amp{} fails to reduce the \gls{ser} throughout the iterations. 
Again, the unperturbed \gls{apsm} settles at a \gls{ser} close to that achieved by the \boxdetector{}.
Superiorization based on the indicator function $\fltwo=\iota_{\setS}$ (\apsmltwo) does not improve the performance compared to the unperturbed basic algorithm (\apsm). By contrast, the \gls{ser} achieved by \apsmlone{} is about an order of magnitude below the unperturbed basic algorithm \apsm{}, even outperforming the more complex \oamp{} detector.

\begin{figure}[ht]
	\centering
	\includegraphics[width=\figurewidth]{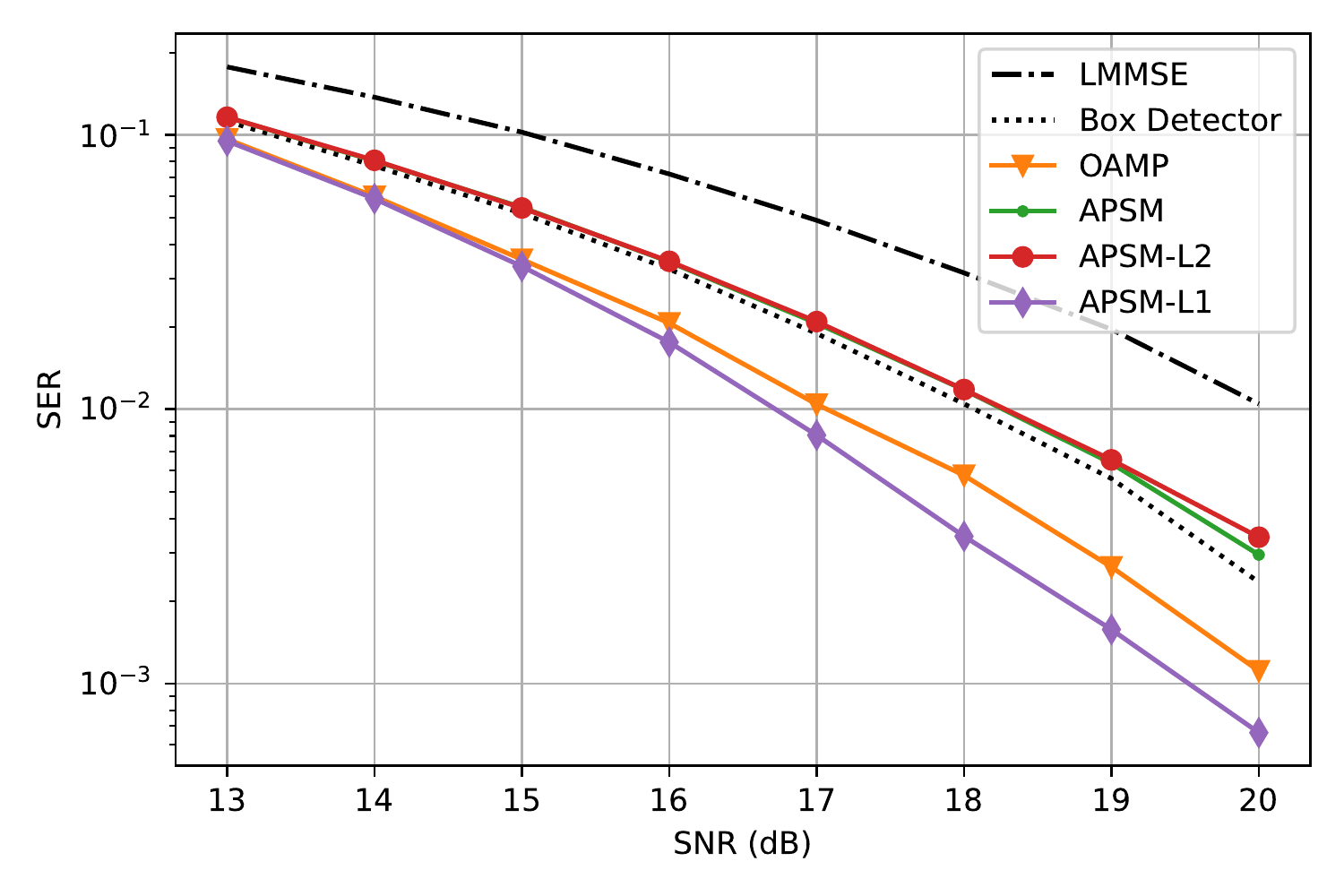}
	\caption{\gls{ser} as a function of the SNR averaged over 10000 3GPP channels.}
	\label{fig:ser_snr_3gpp}
\end{figure}
Figure~\ref{fig:ser_snr_3gpp} shows the average \gls{ser} as a function of the \gls{snr} for 3GPP channels. Since \amp{} did not converge for 3GPP channels, it is excluded from this comparison. 
It can be seen that the perturbations of \apsmltwo{} did not achieve a significant improvement over \apsm{} without perturbations. Both \apsm{} and \apsmltwo{} almost achieve the same \gls{ser} as the \boxdetector{} throughout the entire \gls{snr} range.
By contrast, \apsmlone{} achieves a lower \gls{ser} than \oamp{} for all \gls{snr} levels.

%%% Conclusion %%%%%%%%%%%%%%%%%%%%%%%%%%%%%%%%%%%%%%%%%%%%%%%%%%%%%%%%%%%%%%%%

\section{Conclusion}
In this paper, we derived conditions for the convergence of perturbed versions of the \gls{apsm} and we proposed iterative \gls{mimo} detectors with convergence guarantees based on a superiorized \gls{apsm}. 
Unlike \gls{iolama}, the proposed methods are guaranteed to converge for arbitrary channel matrices.
Simulations show that the proposed methods can outperform \gls{oamp} on realistic channels. Moreover, in contrast to \gls{oamp}, the proposed detectors do not require matrix inverses, so they have a per-iteration complexity similar to \gls{iolama}.
The theoretical results in Section~\ref{sec:apsm_bpr} are valid in arbitrary (possibly infinite dimensional) real Hilbert spaces. Owing to the wide applicability of the \gls{apsm}, they can be used to devise heuristics with convergence guarantees for various other applications.

\section{Appendix}
\begin{appendices}
\subsection{Proof of Proposition~\ref{prop:qne_qf_type1}}\label{apx:qne_qf_type1}
By quasi-nonexpansivity of $T_n$, it holds that $(\forall \z\in\fixInter)(\forall n\in\NN)$
\begin{small}
\begin{align*}
    \|\x_{n+1} - \z\|^2 
    &= \| T_n(\x_n + \beta_n\y_n) - \z\|^2\\
    &\overset{(i)}{\le} \| \x_n + \beta_n\y_n - \z\|^2\\
    &= \| \x_n - \z\|^2 + 2\beta_n\langle\x_n - \z, \y_n\rangle + \beta_n^2\|\y_n\|^2\\
    &\overset{(ii)}{\le} \| \x_n - \z\|^2 + 2\beta_n\|\x_n - \z\|\cdot\|\y_n\| + \beta_n^2\|\y_n\|^2\\
    &= \left(\| \x_n - \z\| + \beta_n\|\y_n\|\right)^2,
\end{align*}
\end{small}
where (i) follows from quasi-nonexpansivity of $T_n$ and (ii) is an application of the Cauchy-Schwartz inequality.
Since $(\beta_n\y_n)_{n\in\NN}$ is a sequence of bounded perturbations, there exists $r>0$ such that $(\forall n\in\NN)$ $\|\y_n\|\le r$, and $(\gamma_n)_{n\in\NN}:=(r\beta_n)_{n\in\NN}\in\lplusone$.
Consequently, $\left(\exists (\gamma_n)_{n\in\NN}\in\lplusone\right)$ $(\forall \z\in\fixInter)$ $(\forall n\in\NN)$
\begin{equation*}
    \|\x_{n+1} - \z\| \le \|\x_n - \z\| + \gamma_n,
\end{equation*}
which is the desired result. \pushQED{\qed}\qedhere\popQED

\subsection{Proof of Proposition~\ref{prop:kappa-attracting_quasi-fejer}}\label{apx:kappa-attracting_quasi-fejer}
Since $(\forall n\in\NN)$ $T_n$ is $\kappa$-attracting quasi-nonexpansive with $\fix(T_n)\supseteq\fixInter\supseteq\setU$, it holds that $(\forall \z\in\setU) (\forall n\in\NN)$
% \begin{small}
\begin{align*}
     &\|\x_{n+1} - \z\|^2\\
     &\le \|\x_{n} + \beta_n\y_n - \z\|^2 -  \kappa\|\x_{n+1} - (\x_n + \beta_n\y_n)\|^2\\
    %  &=  \|\x_{n} + \beta_n\y_n - \z\|^2 -  \kappa\left(\|\x_{n+1} - \x_n\|^2 + 2\beta_n\langle\y_n,\x_n - \x_{n+1}\rangle + \beta_n^2\|\y_n\|^2\right)\\
    &= \|\x_{n} - \z\|^2 + 2\beta_n\langle \x_{n} - \z , \y_n\rangle + \beta_n^2\|\y_n\|^2 \\
    &\quad -  \kappa\left(\|\x_{n+1} - \x_n\|^2 + 2\beta_n\langle\y_n,\x_n - \x_{n+1}\rangle + \beta_n^2\|\y_n\|^2\right)\\
    &\overset{(i)}{\le} \|\x_{n} - \z\|^2 + 2\beta_n\langle \x_{n} - \z , \y_n\rangle + \beta_n^2\|\y_n\|^2 \\
    &\quad -  \kappa\left(\|\x_{n+1} - \x_n\|^2 + 2\beta_n\langle\y_n,\x_n - \x_{n+1}\rangle \right)\\
    &\overset{(ii)}{\le} \|\x_{n} - \z\|^2 + 2\beta_n\|\x_{n} - \z\| \|\y_n\| + \beta_n^2\|\y_n\|^2 \\
    &\quad -  \kappa\|\x_{n+1} - \x_n\|^2 + 2\kappa\beta_n\|\y_n\| \|\x_n - \x_{n+1}\| \\
    &= \|\x_{n} - \z\|^2 -  \kappa\|\x_{n+1} - \x_n\|^2\\
    &\quad + 2\beta_n \|\y_n\|\left(\|\x_{n} - \z\| + \kappa\|\x_n-\x_{n+1}\|\right) + \beta_n^2\|\y_n\|^2,
\end{align*}
% \end{small}
where (i) follows from nonnegativity of $\beta_n\|\y_n\|$ and (ii) is a two-fold application of the Cauchy-Schwartz inequality.
By Proposition~\ref{prop:qne_qf_type1}, $(\x_n)_{n\in\NN}$ is quasi-Fej\'er of Type-I relative to $\fixInter$, so Fact~\ref{fact:qf_bounded} ensures that $(\x_n)_{n\in\NN}$ is bounded.
Boundedness of $(\x_n)_{n\in\NN}$, $(\y_n)_{n\in\NN}$ and $\setU$, guarantee the existence of some $r>0$ such that $(\forall n \in\NN) (\forall \z\in\setU)$ $\|\x_{n} - \z\| + \kappa\|\x_n-\x_{n+1}\|\le r$ and $(\forall n \in\NN)$ $\|\y_n\|\le r$.  Consequently we can write
\begin{align*}
    \|\x_{n+1} - \z\|^2 
    &\le \|\x_n-\z\|^2 - \kappa\|\x_{n+1}-\x_{n}\|^2 +\beta_n r^2\left(2 + \beta_n\right)\\
    &\le \|\x_n-\z\|^2 - \kappa\|\x_{n+1}-\x_{n}\|^2 +\beta_n r^2\left(2 + b\right),
\end{align*}
where we defined $b:=\sum_{n\in\NN}\beta_n$. Therefore, defining $c:=r^2(2+b)$ and $(\gamma_n)_{n\in\NN}:=(c\beta_n)_{n\in\NN}\in\lplusone$ yields the desired result. \pushQED{\qed} \qedhere \popQED

\subsection{Proof of Lemma~\ref{lem:convergence_of_projection}}\label{apx:convergence_of_projection}
The following proof follows line by line the proof in \cite[Lemma~1]{yamada2005adaptive}, while accounting for quasi-Fej\'er monotonicity (instead of Fej\'er monotonicity as in \cite{yamada2005adaptive}).

It is sufficient to show that $(P_\setV(\x_n))_{n\in\NN}$  is a Cauchy sequence. To do so, we first show that there exists $(\gamma_n)_{n\in\NN}\in\lplusone$ such that $(\forall n\in\NN)$
\begin{equation}\label{eq:claim1}
    2\varepsilon \|P_\setV(\u_n)-P_\setV(\u_{n+1})\| \le \|\u_n - \x_0\|^2 - \|\u_{n+1} - \x_0\|^2 + \gamma_n.
\end{equation}
% Since $\setU$ is bounded, (Type-III) quasi-Fejér monotonicity relative to $\setU$ implies quasi-Fejér monotonicity of Type-II relative to $\setU$ (see Fact~\ref{fact:relation_qf_types}).
If $P_\setV(\u_n)= P_\setV(\u_{n+1})$ for some $n\in\NN$, quasi-Fejér monotonicity of $(\u_n)_{n\in\NN}$ ensures that \eqref{eq:claim1} holds for this $n$.
Therefore it is sufficient to consider $n\in\NN$ such that $P_\setV(\u_n)\neq P_\setV(\u_{n+1})$. In this case, we have
$
    \x_0 + \varepsilon\frac{P_\setV(\u_n) - P_\setV(\u_{n+1})}{\|P_\setV(\u_n) - P_\setV(\u_{n+1})\|}\in\setC\cap\setV
$,
thus by Type-I quasi-Fej\'er monotonicity of $(\u_n)_{n\in\NN}$ there exists $(\delta_n)_{n\in\NN}\in\lplusone$ such that
\begin{align*}
    &\left\|\x_0 + \varepsilon\frac{P_\setV(\u_n) - P_\setV(\u_{n+1})}{\|P_\setV(\u_n) - P_\setV(\u_{n+1})\|}- \u_{n+1}\right\|\\
    &\le
    \left\|\x_0 + \varepsilon\frac{P_\setV(\u_n) - P_\setV(\u_{n+1})}{\|P_\setV(\u_n) - P_\setV(\u_{n+1})\|} - \u_n\right\| + \delta_n.
\end{align*}
Squaring and expanding the above inequality yields
\begin{small}
\begin{align*}
    &\|\x_0-\u_{n+1}\|^2 + 2\varepsilon\left\langle
    \frac{P_\setV(\u_n) - P_\setV(\u_{n+1})}{\|P_\setV(\u_n) - P_\setV(\u_{n+1})\|}, \x_0-\u_{n+1}\right\rangle + \varepsilon^2\notag\\
    &\le \|\x_0-\u_n\|^2 + 2\varepsilon\left\langle
    \frac{P_\setV(\u_n) - P_\setV(\u_{n+1})}{\|P_\setV(\u_n) - P_\setV(\u_{n+1})\|}, \x_0-\u_n\right\rangle + \varepsilon^2\notag\\
    & ~ + 2\delta_n \left\|\x_0 + \varepsilon\frac{P_\setV(\u_n) - P_\setV(\u_{n+1})}{\|P_\setV(\u_n) - P_\setV(\u_{n+1})\|} - \u_n\right\| + \delta_n^2.
\end{align*}
\end{small}
By rearranging and applying the triangle inequality, we obtain
\begin{small}
\begin{align*}
    & 2\varepsilon
    \frac{\left\langle P_\setV(\u_n) - P_\setV(\u_{n+1}), \u_n-\u_{n+1}\right\rangle}{\|P_\setV(\u_n) - P_\setV(\u_{n+1})\|}  \notag\\
    &\le \|\x_0-\u_n\|^2 - \|\x_0-\u_{n+1}\|^2\\
    &\quad + \delta_n\left( 2\left\|\x_0 + \varepsilon\frac{P_\setV(\u_n) - P_\setV(\u_{n+1})}{\|P_\setV(\u_n) - P_\setV(\u_{n+1})\|} - \u_n\right\| + \delta_n\right)\notag\\
     &\le \|\x_0-\u_n\|^2 - \|\x_0-\u_{n+1}\|^2 + \delta_n\left( 2\left\|\x_0 - \u_n\right\| +2\varepsilon + \delta_n\right).
\end{align*}
\end{small}
Since  $(\u_n)_{n\in\NN}$ is quasi-Fej\'er monotone with respect to $\setC$ and $\x_0\in\setC$, the sequence $(\|\x_0-\u_n\|)_{n\in\NN}$ converges (see Fact~\ref{fact:qf_bounded}). Therefore $(\exists r>0)$ $(\forall n\in\NN)$ $\left\|\x_0 - \u_n\right\|<r$. By defining  $a:=(2r + 2\varepsilon + \sum_{n\in\NN}\delta_n)$ we obtain a sequence $(\gamma_n)_{n\in\NN} = (a\delta_n)_{n\in\NN}\in\lplusone$ such that
\begin{small}
\begin{align*}
    & 2\varepsilon
    \frac{\left\langle P_\setV(\u_n) - P_\setV(\u_{n+1}), \u_n-\u_{n+1}\right\rangle}{\|P_\setV(\u_n) - P_\setV(\u_{n+1})\|}\\
    &\le \|\x_0-\u_n\|^2 - \|\x_0-\u_{n+1}\|^2 + \gamma_n.
\end{align*}
\end{small}
From firm nonexpansivity of $P_\setV$ (see Definition~\ref{def:nonexpansive_mappings}) we have 
\begin{align*}
0&\le
\|P_\setV(\u_n)-P_\setV(\u_{n+1})\|\\
&\le \frac{\left\langle P_\setV(\u_n) - P_\setV(\u_{n+1}), \u_n-\u_{n+1}\right\rangle}{\|P_\setV(\u_n)-P_\setV(\u_{n+1})\|},
\end{align*}
which proves \eqref{eq:claim1}.
Since $(\forall n\in\NN)$ $\gamma_n\ge0$, the inequality in \eqref{eq:claim1} implies $(\forall n\in\NN)(\forall k\in\NN)$
\begin{align*}
      &2\varepsilon \|P_\setV(\u_n)-P_\setV(\u_{n+k})\| \\
    %  &\le \|\u_n - \x_0\|^2 - \|\u_{n+k} - \x_0\|^2 + \sum_{i=n}^{n+k-1}\gamma_i\notag\\
     &\qquad \le \|\u_n - \x_0\|^2- \|\u_{n+k} - \x_0\|^2 + \sum_{i=n}^{\infty}\gamma_i.
\end{align*}
% Since $(\gamma_n)_{n\in\NN}\in\lplusone$, we can write $\gamma:=\sum_{n\in\NN}\gamma_n$ and
% \begin{equation}
%     \lim_{n\to\infty} \sum_{i=n}^{\infty}\gamma_i=\gamma - \lim_{n\to\infty} \sum_{i=1}^{n-1}\gamma_i = 0.
% \end{equation}
Moreover, since $(\|\u_n-\x_0\|)_{n\in\NN}$ converges and $(\gamma_n)_{n\in\NN}\in\lplusone$, $(\forall \delta>0)$ $(\exists N\in\NN)$ $(\forall n\ge N)$ $(\forall k \in\NN)$ 
\begin{align*}
    & 2\varepsilon \|P_\setV(\u_n)-P_\setV(\u_{n+k})\|\\ &\le\|\u_n - \x_0\|^2 - \|\u_{n+k} - \x_0\|^2 + \sum_{i=n}^{\infty}\gamma_i< \delta,
\end{align*}
\pushQED{\qed}
which shows that $(P_\setV(\u_n))_{n\in\NN}$ is a Cauchy sequence.\qedhere
\popQED

\subsection{Proof of Theorem~\ref{thm:strong_convergence}}\label{apx:strong_convergence}
The following proof follows very closely the proof in \cite{yamada2005adaptive}, extending the result in \cite{yamada2005adaptive} to quasi-Fej\'er monotone sequences. A more detailed proof can be found in \cite[Section~2.2]{fink2022thesis}.

According to Lemma~\ref{lem:convergence_of_projection}, the sequence $(P_\setW(\u_n))_{n\in\NN}$ converges strongly a point in $\setW$. Hence we can define $\vhat:=\lim_{n\to\infty}P_\setW(\u_n)$ and $\e\in\setH$ satisfying $\setW=\{\x\in\setH~|~ \langle\e,\x-\vhat\rangle=0\}$ and $\|\e\|=1$.
Moreover, according to Fact~\ref{fact:qf_bounded}, the sequence $(\|\u_n-\z\|)_{n\in\NN}$ converges for all $\z\in\setC$, so we can define $a:=\lim_{n\to\infty}\|\u_n-P_{\setC\cap\setW}(\vhat)\|$ and $\rho:=\sqrt{a^2-\|P_{\setC\cap\setW}(\vhat)-\vhat\|^2}$. Note that by nonexpansivity of $P_\setW$ we have $a\ge\lim_{n\to\infty}\|P_{\setW}(\u_n)-P_{\setC\cap\setW}(\vhat)\|=\|\vhat-P_{\setC\cap\setW}(\vhat)\|$, so $\rho$ is well-defined.

Now we apply the same geometric arguments as in \cite[Thm.~1]{yamada2005adaptive}, with the slight difference that we replace the set $\setS_{(\delta_1,\delta_2)}:=\{\x\in\setH~|~ \|P_\setW(\x)-\vhat\|\le\delta_1,~ a\le\|\x-P_{\setC\cap\setW}(\vhat)\|\le a+\delta_2\}$
by
$\setS^\prime_{(\delta_1,\delta_2)}:=\{\x\in\setH~|~ \|P_\setW(\x)-\vhat\|\le\delta_1,~ |\|\x-P_{\setC\cap\setW}(\vhat)\|- a|\le\delta_2\}$ to account for quasi-Fej\'er monotonicity (instead of Fej\'er monotonicity) of $(\u_n)_{n\in\NN}$.
As in \cite{yamada2005adaptive}, we deduce the existence of sufficiently small $\delta_1>0$ and $\delta_2>0$ such that $\setS^\prime_{(\delta_1,\delta_2)}\subset\setB(\varepsilon):=\setB_1(\varepsilon)\cup\setB_2(\varepsilon)$, where $\setB_1(\varepsilon):=\{\x\in\setH~|~\|\x-(\vhat+\rho\e)\|\le\varepsilon\}$ and $\setB_2(\varepsilon):=\{\x\in\setH~|~\|\x-(\vhat-\rho\e)\|\le\varepsilon\}$ for arbitrary fixed $\varepsilon\in(0,\rho/2)$.

Since $\lim_{n\to\infty}\|P_\setW(\u_n)-\vhat\|=0$ and $\lim_{n\to\infty}\|\u_n-P_{\setC\cap\setW}(\vhat)\|=a$, there exists $n_1\in\NN$ such that
\begin{equation*}
    (\forall n\ge n_1) \quad \u_n\in\setS^\prime_{(\delta_1,\delta_2)}\subset\setB(\varepsilon).
\end{equation*}
Moreover, by \eqref{eq:thm_conv_quasi_fejer}, there exists $n_2\in\NN$ such that
\begin{equation*}
    (\forall n\ge n_2)\quad \|\u_n - \u_{n+1}\|< 2 \varepsilon,
\end{equation*}
which ensures the unique existence of $i\in\{1,2\}$ satisfying
\begin{equation*}
    (\forall n\ge n_2)\quad \u_n\in\setB_i(\varepsilon).
\end{equation*}
\pushQED{\qed}
This implies the strong convergence of $(\u_n)_{n\in\NN}$ to either $\vhat+\rho\e$ or $\vhat - \rho\e$.\qedhere
\popQED

\subsection{Proof of Theorem~\ref{thm:apsm_bpr}}\label{apx:apsm_bpr}
\begin{enumerate}[(a)]
 \item Note that $(\forall n\in\NN)$ $\lev_{\le 0}\Theta_n\neq\emptyset$ by assumption. Hence by Fact~\ref{fact:sg_proj_attracting}, $(\forall n\in\NN)$ the mapping $T_n$ is quasi-nonexpansive. According to Proposition~\ref{prop:qne_qf_type1}, the sequence $(\x_n)_{n \in\NN}$ is quasi-Fejér monotone of Type-I relative to $\apsmInter$. This in turn implies that $(\x_n)_{n\in\NN}$ is bounded (see Fact~\ref{fact:qf_bounded}).
 
 \item   Introducing the shorthand $(\forall n\in\NN)$ $\z_n:=\x_n + \beta_n\y_n$ and
     \begin{equation}\label{eq:phin_shorthand}
    \phin = \begin{cases}
     \lambda_n\frac{\Theta_n(\z_n)}{\|\Theta_n^\prime( \z_n)\|^2}\Theta_n^\prime(\z_n) & \text{if $\Theta_n^\prime(\z_n)\neq \Null$}, \\
    \Null & \text{otherwise}
    \end{cases}
    \end{equation}
    % we can write $(\forall n\ge N_0)$
    we can write $(\forall n\in\NN)$
    $(\forall \x_n\in\setK)$  $(\forall\z\in\apsmInter)$
    \begin{align*}
        \|\x_{n+1}-\z\|^2 &= \| P_\setK(\z_n - \phin) - P_\setK(\z)\|^2\\
        % &= \| P_\setK(\x_n + \beta_n\y_n - \phin) - P_\setK(\z)\|^2\\
        &\overset{(i)}{\le} \|\x_n + \beta_n\y_n - \phin - \z\|^2\\
        % &= \|\x_n - \z\|^2 - 2\langle \phin, \x_n - \z\rangle \\
        % &\quad + 2\beta_n\langle\y_n,\x_n - \z\rangle + \|\beta_n\y_n - \phin\|^2\\
        &= \|\x_n - \z\|^2  + 2\langle\beta_n\y_n - \phin,\x_n - \z\rangle\\
        &\quad + \|\beta_n\y_n - \phin\|^2\\
        % &= \|\x_n - \z\|^2 - 2\langle \phin, \x_n - \z\rangle + 2\beta_n\langle\y_n,\x_n - \z\rangle \\
        % &\quad + \beta_n^2\|\y_n\|^2 -2\langle \phin,\beta_n\y_n\rangle + \|\phin\|^2\\
        &= \|\x_n - \z\|^2 + 2\langle\beta_n\y_n - \phin,\x_n - \z\rangle\\
        &\quad + \beta_n^2\|\y_n\|^2 -2\langle \phin,\beta_n\y_n\rangle + \|\phin\|^2\\
        &= \|\x_n - \z\|^2 - 2\langle \phin, \x_n + \beta_n\y_n- \z\rangle \\
        &\quad + 2\beta_n\langle\y_n,\x_n - \z\rangle + \beta_n^2\|\y_n\|^2  + \|\phin\|^2\\
        &\overset{(ii)}{\le}  \|\x_n - \z\|^2 - 2\langle \phin, \z_n- \z\rangle  + \|\phin\|^2\\
        &\quad + 2\beta_n\|\y_n\|\|\x_n - \z\| + \beta_n^2\|\y_n\|^2 
    \end{align*}
    where (i) follows from nonexpansivity of $P_\setK$, and (ii) is an application of the Cauchy-Schwarz inequality.
    % , and (iii) follows from Fact~\ref{fact:subgradient_projection}, because the mapping $\x_n\mapsto\x_n+\phin$ is a relaxed subgradient projector relative to $\Theta_n$ with relaxation parameter $\lambda_n\in[0,2]$.
    Since $(\x_n)_{n\in\NN}$ and $(\y_n)_{n\in\NN}$ are bounded, for any bounded subset $\setU\subset\apsmInter$ there exists $r>0$ such that $(\forall \z\in\setU)$ $(\forall n\in\NN)$ $\|\x_n-\z\|\le r$ and $\|\y_n\|\le r$. Hence by defining $c:=(2r^2 + r^2\sum_{n\in\NN}\beta_n)$ and $(\gamma_n)_{n\in\NN}:=(c\beta_n)_{n\in\NN}$ we have $(\forall\z\in\setU)$
    \begin{align*}
         \|\x_{n+1}-\z\|^2
         &\le \|\x_n  - \z\|^2 -  2\left\langle\phin, \z_n - \z\right\rangle\\
         &\quad + \|\phin\|^2
         + 2\beta_{n}r^2 +  \beta_{n}^2r^2\\
     &\le \|\x_n  - \z\|^2 -  2\left\langle\phin, \z_n  - \z\right\rangle\\
     &\quad + \|\phin\|^2 + \gamma_n.
    \end{align*}

    If $\Theta_n(\z_n)=0$ or $\Theta_n^\prime( \z_n)=\Null$, \eqref{eq:phin_shorthand} yields $\phin=\Null$, whereby
    \begin{equation}\label{eq:boundedness_proof1}
             \|\x_{n+1}-\z\|^2 \le \|\x_n  - \z\|^2 +  \gamma_n.
    \end{equation}
    Otherwise, i.e., if $\Theta_n(\z_n)\neq0$ and $\Theta_n^\prime(\z_n)\neq\Null$, it follows from \eqref{eq:subdifferential} that
    \begin{align}\label{eq:boundedness_proof2}
                &~ \|\x_{n+1}-\z\|^2 \notag\\
                &\le  \|\x_n  - \z\|^2 -  2\lambda_n\frac{\Theta_n( \z_n)}{\|\Theta_n^\prime( \z_n)\|^2}\left\langle\Theta_n^\prime( \z_n), \z_n  - \z\right\rangle \notag\\
     &\quad + \lambda_n^2\frac{\Theta_n(\z_n)^2}{\|\Theta_n^\prime(\z_n)\|^2} +  \gamma_n\notag\\
     &\le \|\x_n  - \z\|^2 -  2\lambda_n\frac{\Theta_n(\z_n)}{\|\Theta_n^\prime(\z_n)\|^2}\left(\Theta_n(\z_n)-\Theta_n(\z)\right) \notag\\
     &\quad + \lambda_n^2\frac{\Theta_n(\z_n)^2}{\|\Theta_n^\prime(\z_n)\|^2} +  \gamma_n\notag\\
     &=  \|\x_n  - \z\|^2 \notag\\
     &\quad -\lambda_n\left(2\left(1-\frac{\Theta_n(\z)}{\Theta_n(\z_n)}\right)-\lambda_n\right)\frac{\Theta_n^2(\z_n)}{\|\Theta_n^\prime(\z_n)\|^2} +  \gamma_n.
    \end{align} 
    
    Since $(\forall n\in\NN)$ $\lambda_n\in[\varepsilon_1,2-\varepsilon_2]$ and
    % $(\forall n> N_0)$
    $(\forall n\in\NN)$
    $(\forall \z\in\setU\subset\apsmInter)$ $\Theta_n(\z)=0$, we have
    \begin{equation*}
        \lambda_n\left(2\left(1-\frac{\Theta_n(\z)}{\Theta_n(\z_n)}\right)-\lambda_n\right)
        \ge
    \varepsilon_1\varepsilon_2.
    \end{equation*}
   Hence, by defining a sequence
    % \begin{equation}\label{eq:aux_sequence}
    %     (\forall n\in\NN)\quad c_n:= \begin{cases}
    %     0 & \text{if $\Theta_n^\prime(\x_n)=\Null$}\\
    %     \lambda_n\left(2\left(1-\frac{\Theta_n(\z)}{\Theta_n(\x_n)}\right)-\lambda_n\right)\frac{\Theta_n^2(\x_n)}{\|\Theta_n^\prime(\x_n)\|^2} & \text{otherwise},
    %     \end{cases}
    % \end{equation}
    \begin{equation}\label{eq:aux_sequence2}
        (\forall n\in\NN)\quad  c_n:= \begin{cases}
        0 & \text{if $\Theta_n^\prime(\z_n)=\Null$}\\
        \varepsilon_1\varepsilon_2\frac{\Theta_n^2(\z_n)}{\|\Theta_n^\prime(\z_n)\|^2} & \text{otherwise},
        \end{cases}
    \end{equation}
    we can summarize \eqref{eq:boundedness_proof1} and \eqref{eq:boundedness_proof2} as
    \begin{equation}
        (\forall n \in\NN)\quad \|\x_{n+1}-\z\|^2 \le  \|\x_n  - \z\|^2 -c_n +  \gamma_n.
    \end{equation}
    Because \eqref{eq:subdifferential} ensures that $\Theta_n^\prime(\x)=\Null\Rightarrow \Theta_n(\x)=\Theta_n^\star=0$, it is sufficient to consider the case $\Theta_n^\prime(\z_n)\neq\Null$.
    Moreover, if this case occurs finitely many times, there exists $N_0$ such that $(\forall n\ge N_0)$ $\Theta_n^\star=0$.
    Thus it remains to show that $\lim_{k\to\infty}\Theta_{n_k}( \z_{n_k})=0$, where $(n_k)_{k\in\NN}$ is the subsequence
    of $(n)_{n\in\NN}$ comprised of all elements of the infinite set
    $\setJ:=\{n\in\NN~|~\Theta_n^\prime(\x)\neq\Null\}$.
  According to Fact~\ref{fact:quasi_fejer}, 
   $(\|\x_n-\z\|^2)_{n\in\NN}$
   converges and $\left(c_n\right)_{n\in\NN}$ is summable, whereby  
%   $\sum_{n\ge N_0}\emin\emax\frac{\Theta_n^2(\z_n)}{\|\Theta_n^\prime(\z_n)\|^2}<\infty$.
   $\sum_{n\in\setJ}\emin\emax\frac{\Theta_n^2(\z_n)}{\|\Theta_n^\prime(\z_n)\|^2}<\infty$.
   Moreover, since $(\forall n\in\setJ)$ $\emin\emax\frac{\Theta_n^2(\z_n)}{\|\Theta_n^\prime(\z_n)\|^2}\ge0$, it follows that
\begin{equation*}
    \underset{k\to\infty}{\lim} \emin\emax\frac{\Theta_{n_k}^2( \z_{n_k})}{\|\Theta_{n_k}^\prime( \z_{n_k})\|^2}=0.
\end{equation*}
Therefore, boundedness of $\left(\Theta_n^\prime(\x_n+\beta_n\y_n)\right)_{n\in\NN}$ ensures that $\lim_{n\to\infty}\Theta_n(x_n+\beta_n\y_n)=0$.
    
%     Because \eqref{eq:subdifferential} ensures that $\Theta_n^\prime(\x)=\Null\Rightarrow \Theta_n(\x)=\Theta_n^\star=0$, it is sufficient to consider the case $\Theta_n^\prime(\x_n)\neq\Null$.
%   According to Fact~\ref{fact:quasi_fejer}, 
%   $(\|\x_n-\z\|^2)_{n\in\NN}$
%   converges and 
% %   $\left(\hat c_n\right)_{n\ge N_0}$
%   $\left(\hat c_n\right)_{n\in\NN}$
%   is summable, i.e.,  
% %   $\sum_{n\ge N_0}\emin\emax\frac{\Theta_n^2(\z_n)}{\|\Theta_n^\prime(\z_n)\|^2}<\infty$.
%   $\sum_{n\in\NN}\emin\emax\frac{\Theta_n^2(\z_n)}{\|\Theta_n^\prime(\z_n)\|^2}<\infty$.
%   Moreover, since $(\forall n\in\NN)$ $\emin\emax\frac{\Theta_n^2(\z_n)}{\|\Theta_n^\prime(\z_n)\|^2}\ge0$, it follows that
% \begin{equation*}
%     \underset{n\to\infty}{\lim} \emin\emax\frac{\Theta_n^2(\z_n)}{\|\Theta_n^\prime(\z_n)\|^2}=0.
% \end{equation*}
% Therefore, boundedness of $\left(\Theta_n^\prime(\z_n)\right)_{n\in\NN}$ ensures $\lim_{n\to\infty}\Theta_n(\z_n)=0$. \qedhere
% \popQED

 \item It holds by assumption in \eqref{eq:conditions1} that $(\forall n\in\NN)$ $\setK\cap\lev_{\le 0}\Theta_n\neq\emptyset$. Therefore, according to Fact~\ref{fact:sg_proj_attracting}, the mapping $T_n$ in \eqref{eq:apsm_mapping} is $\left(1-\frac{\lambda_n}{2}\right)$-attracting quasi-nonexpansive.
 Since the set $\setU$ is bounded and $(\forall n\in\NN)$ $\lambda_n\le 2-\varepsilon_2$, Proposition~\ref{prop:kappa-attracting_quasi-fejer} implies that $(\exists (\gamma_n)_{n\in\NN}\in\lplusone) (\forall\z\in\setU)$
 \begin{equation*}
    \frac{\varepsilon_2}{2}\|\x_{n+1}-\x_n\|^2\le\|\x_n-\z\|^2 -\|\x_{n+1}-\z\|^2 + \gamma_n.
 \end{equation*}
% Furthermore, since by (a) $(\x_n)_{n\in\NN}$ is a quasi-Fejér sequence of Type-I relative to $\apsmInter$,
Consequently, since $(\x_n)_{n\in\NN}$ is quasi-Fej\'er of Type-I relative to $\apsmInter$ (see (a)), Theorem~\ref{thm:strong_convergence} guarantees that the sequence $(\x_n)_{n\in\NN}$ converges strongly to a point $\hat \u\in \setH$.
More precisely, it holds that $\hat\u\in\setK$, because $(\x_n)_{n\in\NN}$ is a sequence in the closed set $\setK$.
Since $(\beta_n\y_n)_{n\in\NN}$ are bounded perturbations, the sequence $(\z_n:=\x_n+\beta_n\y_n)_{n\in\NN}$ satisfies $\lim_{n\to\infty}\z_n=\lim_{n\to\infty}\x_n=\hat\u$.
By assumption (ii) there exists $R>0$ such that $(\forall n\in\NN)$ $\|\Theta_n^\prime(\hat\u)\|\le R$. Thus
% , defining $(\forall n\in\NN)$ $\z_n=\x_n+\beta_n\y_n$, 
we can use Fact~\ref{fact:subdifferential} and the Cauchy-Schwartz inequality to obtain
\begin{align*}
    0 &\le \Theta_n(\hat\u)\le \Theta_n(\z_n) - \left\langle\z_n-\hat\u,\Theta_n^\prime(\hat\u)\right\rangle\\
    &\le \Theta_n(\z_n) + R\|\z_n-\hat\u\| \to 0.
\end{align*}
%The remainder of this proof is identical to \cite[Thm.~2(c)]{yamada2005adaptive}. 

\item 
The proof is identical to  the proof of \cite[Thm.~2(d)]{yamada2005adaptive}. \pushQED{\qed}\qedhere\popQED
% \pushQED{\qed}\qedhere
% \popQED
 \end{enumerate}
\end{appendices}

\bibliographystyle{IEEEtran}
\bibliography{IEEEabrv, refs_abrv}	
\end{document}